\documentclass[aps,prd,reprint,showpacs,nofootinbib,twoside,notitlepage]{revtex4-1} 
\usepackage{graphicx}
\usepackage{euscript,amssymb}
\usepackage{amsfonts}
\usepackage{amssymb}
\usepackage{amsbsy}
\usepackage{amsmath}
\usepackage{nccmath}   % for de-centering equations environments  i.e.  fleqn environment within which equations envionments are written
\usepackage{color}
\usepackage{bbold}
\usepackage{hyperref}
\usepackage{physics}
\usepackage[caption=false]{subfig} %allows multiple plots in the same figure

\newcommand{\be}{\begin{equation}}
  \newcommand{\ee}{\end{equation}}
\newcommand{\ben}{\begin{eqnarray*}}
  \newcommand{\een}{\end{eqnarray*}}
\newcommand{\bea}{\begin{eqnarray}}
  \newcommand{\eea}{\end{eqnarray}}
\newcommand{\bdm}{\begin{displaymath}}
  \newcommand{\edm}{\end{displaymath}}
\newcommand{\ba}{\begin{align}}
  \newcommand{\ea}{\end{align}}

\begin{document}

\title{From classical to quantum Oppenheimer-Snyder model: non-marginal case} 

\author{Claus Kiefer}

\email{kiefer@thp.uni-koeln.de}

\author{Hamid Mohaddes}

\email{s6hamoha@uni-bonn.de}

\affiliation{University of Cologne, Faculty of Mathematics and Natural Sciences, Institute for Theoretical Physics, Cologne, Germany}

\date{\today}

\begin{abstract}

We first present a consistent canonical formulation of the general (non-marginal) Oppenheimer-Snyder model. The switching between comoving and stationary observer is achieved by promoting coordinate transformations between dust proper time and Schwarzschild-Killing time to canonical ones.
This leads to a multivalued Hamiltonian which is deparameterizable. We then discuss the quantization of comoving and stationary observers by employing the method of Affine Coherent State Quantization (ACSQ). We thereby demonstrate that under certain conditions the quantum corrected trajectories can replace the classical singularity by a bounce. We then show that both comoving and stationary observers see this bouncing collapse behavior. We finally discuss a switching between these classes of observers at the quantum level.

\end{abstract}

\maketitle

%%%%%%%%%%%%%%%%%%%%%%%%%%%%%%%%%%%%%%%%%%%%%
%%%%%%%%%%%%%%% INTRODUCTION %%%%%%%%%%%%%%%%
%%%%%%%%%%%%%%%%%%%%%%%%%%%%%%%%%%%%%%%%%%%%%
\section{Introduction}
\label{introduction}
One expects that a consistent quantum theory of gravity is free of the space-time singularities that plague the classical theory of general relativity \cite{Kiefer:2004xyv}. This has particular relevance for cosmology and black holes. As long as a full theory of quantum gravity is not yet available, one has to resort to particular approaches and, in general, to particular models with a high degree of symmetry. For this purpose, many investigations employ the framework of canonical quantization with metric variables, with the Wheeler-DeWitt equation for the quantum-geometrodynamical wave function(al) as its central equation \cite{Kiefer:2004xyv,PhysRev.160.1113}. This is a very conservative approach within which important conceptual questions can be posed and answered, even if this approach will not turn out to be the final one at all scales. It is also the approach that we shall use in our paper. 

Concerning singularity avoidance, already DeWitt has formulated a heuristic criterion stating that the wave function should vanish in the regions of classical singularities \cite{PhysRev.160.1113}. This criterion can be generalized to accommodate the conformal nature of configuration space \cite{Kiefer:2019bxk} and has found widespread applications in cosmology, in particular for singularities that occur in classical models with dark energy; see, for example, \cite{Bouhmadi-Lopez:2019zvz} and the references therein. Since the universe is approximately homogeneous, there are no asymptotic regions in cosmology. In contrast to this, for situations involving (isolated) black holes, there exists an asymptotically flat region which can serve as a reference frame. One can, in particular, demand that the quantum behavior is unitary with respect to the asymptotic (semiclassical) time. By this unitarity, wave packets cannot just ``disappear'' during the quantum evolution, and thus unitarity can be used as a criterion for singularity avoidance. This is, of course, reminiscent of unitarity as a solution to the black-hole information problem. 

Examples of singularity avoidance for black holes have been widely discussed in the literature, e.g. for the collapse of a dust shell \cite{Hajicek:2001yd} or for the collapse of a dust cloud in the Lema\^{\i}tre-Tolman-Bondi (LTB) model \cite{LTB}; see \cite{Malafarina:2017csn} for a review. A model of particular interest is the Oppenheimer-Snyder (OS)-model of homogeneous dust collapse. This was, in fact, historically the first analytic solution in general relativity describing the collapse of an object to a black hole. In this model, the dust cloud is described by a Friedmann-Lema\^{\i}tre-Robertson-Walker (FLRW) model, while the exterior of the dust cloud is represented by the Schwarzschild metric. 
Classical and quantum aspects of the OS-model were recently discussed in \cite{tim} \cite{tim1}, where it was shown in which sense and in which situations the classical singularity can be avoided in the quantum theory: the classical trajectories ending at a singularity are replaced by wave packets exhibiting a {\em bounce}. In these papers, however, restriction was made to the case of a {\em flat} (marginal) FLRW-model, corresponding to the choice $k=0$ for the curvature parameter. The main purpose of the present paper is the generalization to the non-marginal case. Only then  we can consider the analysis of the quantum OS-model as complete. We shall, in fact, find that singularity avoidance can still be achieved. We can thus consider the quantum OS-model as a consistent model of quantum gravity in the geometrodynamic approach. 

Our paper is organized as follows. In sections \ref{ch:chapter_2} and \ref{ch:chapter_3} we discuss the canonical OS-model and  apply to it the Kucha\v{r} decomposition. This leads to a boundary term which renders the action non-canonical. In order to bring the action into canonical form, we introduce in sections \ref{chapter 4} and \ref{chapter 5} different coordinate systems. With these, the problematic boundary term can be avoided. In section \ref{static observer hamiltonian 6}, the Hamiltonian for the stationary observer is derived. Due to the unusual form of the Hamiltonian, see our equation \eqref{stationary hamiltonian finale 4} below,  the standard Dirac quantization program cannot be employed. We thus introduce in section \ref{chapter 7} another quantization scheme known as Affine Coherent State Quantization (ACSQ); see, for example \cite{tim} and \cite{almeida2018three} for details on ACSQ. We then apply this method in sections \ref{comoving obsever qm theory} and \ref{quantum theory for the stationary observer} to the comoving and stationary observers, respectively, and discuss the results. In section \ref{switching 10}  we then address the switching between these two observers. In section \ref{summary} we present our summary and conclusions.
Some technical details are relegated to four Appendices.

%%%%%%%%%%%%%%%%%%%%%%%%%%%%%%%%%%%%%%%%%%%%%%%%%%%%%%%%%%%%%%%%%%

\section{Canonical Oppenheimer-Snyder model}
\label{ch:chapter_2}
In this section, we shall briefly introduce and discuss the canonical Oppenheimer-Snyder (OS) model. We start with the ADM-decomposed action 
for spherically symmetric gravity with Brown-Kucha\v{r} dust \cite{Brown:1994py}. The full action reads
\begin{equation}
    \begin{aligned}
\label{action 4.0}
S=&\int dt \int_0^{\infty} dr\, \left(P_{\tau}\dot{\tau}+P_R\dot{R}+ P_{\Lambda}\dot{\Lambda}-NH-N^rH_r\right)\\- &\int dt\, M_+ \dot{T}_+,
\end{aligned}
\end{equation}
with
\begin{align}
\label{constraint 4.1}
    H&=\frac{\Lambda}{2R^2}P_{\Lambda}^2-\frac{P_RP_{\Lambda}}{R}+\frac{RR''}{\Lambda}-\frac{\Lambda'R'R}{\Lambda^2}+\frac{R'^2}{2\Lambda}\\- \nonumber&\frac{\Lambda}{2}+P_{\tau}\sqrt{1+\frac{\tau'^2}{\Lambda^2}},\\
    H_r&=P_RR'-P_{\Lambda}\Lambda'+P_{\tau}\tau',
    \label{constraint 4.2}
\end{align}
where $\tau$ is the dust proper time, $T^{+}$ the Schwarzschild Killing time at right infinity,
and $M^+$ the ADM mass of space-time; $\Lambda(r,t),R(r,t)$ are components of a spherically symmetric metric on the leaves of foliation with label time $t$,
\begin{equation}
    d\sigma^2=\Lambda^2 dr^2+ R^2 d\Omega^2.
    \label{spherically symmetric metric}
\end{equation}
The boundary term in \eqref{action 4.0} will be central to the development of the canonical OS-model.
The action above describes any spherically symmetric space-time generated by Brown-Kucha\v{r} non-rotating dust. Our goal here is to implement the OS-model into the above Hamiltonian. We recall that the surface of the dust cloud in the OS-model is at coordinate radius $r_s>0$, and outside of the cloud the mass density $\rho$ vanishes. We can implement this using the canonical variable $P_{\tau}$,
\begin{equation}
    P_{\tau}=4\pi \Lambda^2 R^2\sqrt{1+\frac{\tau'^2}{\Lambda^2}}\rho.
    \label{dust4}
\end{equation}
In addition, we know that the OS-model consists of pressureless homogeneous dust; therefore the ADM foliation $r\leq r_s$ should coincide with hypersurfaces of constant $\rho,\tau$. The corresponding FLRW-metric should be of the form
\begin{equation}
\label{flrw 4}
    dS^2=-N^2dt^2+a^2(t)\left(\frac{dr^2}{1-kr^2}+r^2 d\Omega^2\right),
\end{equation}
where $k$ assumes the values $(0,\pm 1)$ for flat, open, and closed sections of the FLRW-space-time. We implement furthermore the restrictions 
\begin{align*}
    N^{r}&=0,\\
    N&=\Bar{N}.
\end{align*}
Comparing \eqref{spherically symmetric metric} and \eqref{flrw 4}, one arrives at 
\begin{align}
\label{match 4.1}
    \Lambda&=\frac{a}{\sqrt{1-kr^2}},\\
    R&=ar.
    \label{match 4.2}
\end{align}
Rewriting $P_{\tau}$, \eqref{dust4}, using \eqref{match 4.1} and \eqref{match 4.2} and considering hypersurfaces of constant $\rho,\tau$, we have
\begin{equation}
    P_{\tau}=\frac{4\pi r^2 a^3}{\sqrt{1-kr^2}}\equiv \frac{r^2}{V_s\sqrt{1-kr^2}}\Bar{P_{\tau}},
    \label{P tau}
\end{equation}
with  
\begin{equation}
    V_s=\int_0^{r_s} dr \frac{r^2}{\sqrt{1-kr^2}}\,.
    \label{Vs}
\end{equation}
In addition, calculating $P_R,P_{\Lambda}$ using \eqref{match 4.1} \eqref{match 4.2}, we get

\begin{align}
\label{4,p lambda}
  P_{\Lambda}&=-\frac{R}{N}\left(\Dot{R}-R'N^r\right)=-\frac{a\Dot{a}r^2}{N},\\ \label{4 PR}
  P_R&=-\frac{\Lambda}{N}\left(\dot{R}-R'N^r\right)-\frac{R}{N}\left(\Dot{\Lambda}-(\Lambda N^r)'\right)\\ \nonumber &=-\frac{2a\dot{a}r}{N\sqrt{1-kr^2}}.
  \end{align}
   Using \eqref{match 4.1}, \eqref{match 4.2}, \eqref{P tau} , \eqref{4,p lambda}, and \eqref{4 PR} and rewriting the Liouville form, we find 
   
\begin{equation}
\begin{aligned}
    \int_0^{r_s} dr (P_{\tau}\Dot{\tau}+&P_R\Dot{R}+P_{\Lambda}\Dot{\Lambda})=\\ &\int_0^{r_s} dr \left(P_{\tau}\Dot{\tau}-\frac{3a\Dot{a}^2r^2}{\Bar{N}\sqrt{1-kr^2}}\right),
    \end{aligned}
\end{equation}
where $P_{\tau} $ is the conjugate momentum to $\tau$, while the conjugate momentum to $a$ reads
\begin{equation}
    P=\frac{-3V_s a\Dot{a}}{N}.
    \label{P a}
\end{equation}
Using \eqref{P a}, we rewrite \eqref{4,p lambda} and \eqref{4 PR}; inserting these equations into \eqref{constraint 4.1} and \eqref{constraint 4.2}, we get the Hamiltonian for $r\leq r_s$,
\begin{equation}
   \Bar{H}= \int_0^{r_s}\left(NH+N^rH_r\right)=\Bar{N}\left(-\frac{P^2}{6V_s a}-\frac{3V_s}{2}ka+P_{\tau}\right).
   \label{Hamiltonian interior 4}
\end{equation}
The full action can then be formulated by adding the actions of interior (FLRW) and exterior (Schwarzschild) metrics. It reads
\begin{equation}
\begin{aligned}
   S= &\int dt \left(P\Dot{a}+P_{\tau}\Dot{\tau}-\Bar{N}\Bar{H}\right)\\&+ \int dt \int dr \left(P_R\Dot{R}+P_{\Lambda}\Dot{\Lambda}-NH-N^rH_r\right)& \\ &-\int dt M_+\Dot{T}_+ .
    \end{aligned}
    \label{action 4.finale.1}
\end{equation}
The expression $NH+N^rH_r$ is the Hamiltonian for the exterior and $N\Bar{H}$ the expression for the interior Hamiltonian of the OS-model. Although a similar analysis can be found in \cite{Casadio:1998ta}, here we stay close to the formalism presented in \cite{tim1}  which heavily employs the series of canonical transformations discussed in the process of Kucha\v{r}-decomposition \cite{Kuchar:1994zk} . In summary, we have derived the Hamiltonian for the interior region $r<r_s$ given by equation \eqref{Hamiltonian interior 4}, where the dust density does not vanish. The dust density is homogeneous and thus the space-time metric is of FLRW-form. This allowed us to integrate out radial degrees of freedom, leading to the canonical formulation of a Friedmann model with dust as matter, which is described by the scale factor $a$ and its momentum $p_a$ as well as dust proper time $\tau$ and its momentum $P_{\tau}$. We identify $M=(r_s^3/3V_s)P_{\tau}$ as the total mass $M>0$ of the collapsing body. The Hamiltonian constraint in the interior can then be expressed in the form $H+P_{\tau}\approx 0$, which is deparameterizable with respect to dust proper time. Note that the Hamiltonian is always negative, but this does not pose any issues because we identify $H=-(3V_s/r_s^3 \sqrt{p_k})M$; therefore, the Hamiltonian is negative, while the energy is positive.

%%%%%%%%%%%%%%%%%%%%%%%%%%%%%%%%%%%%%%%%%%%%%%%%%%%%%%%%%%%%%%%%%%%%%%%%%
\section{Kucha\v{r} decomposition for the OS-model} \label{ch:chapter_3}

A glance at the constraints \eqref{constraint 4.1} and \eqref{constraint 4.2} reveals that Dirac quantization is very difficult if not impossible. In \cite{Kuchar:1994zk}, Kucha\v{r} developed a method later called Kucha\v{r} decomposition. This is one the most important tools in order to simplify and to quantize the canonical OS-model. The idea is as follows: one performs a series of elaborate canonical transformations and splits up the canonical variables into truly physical degrees of freedom (Dirac observables) and embedding variables (telling us how the leaves of foliation serving as slices of equal time are located in full four dimensional space-time).
The momenta conjugated to these variables should vanish. This method allows one to remove the embedding from the system; only the true physical degrees of freedom remain and only these are considered for quantization. The disadvantage of this method is that it only works in particular (typically highly-symmetric) situations \cite{Kiefer:2004xyv}. 

To apply this method  here, the space-time variables $T,R$ of the Schwarzschild space-time are elevated to canonical coordinates $T(r),R(r)$ on the phase space of spherically symmetric space-times.
The whole dynamical content of the Hamiltonian theory is reduced to constraints 
requiring $P_T(r),P_R(r)$ to vanish. This simplifies the problem of quantization and turns a hopeless problem to an easily soluble one; for details we refer to \cite{Kuchar:1994zk}.

In this subsection, we apply the Kucha\v{r} decomposition method to the OS-model. We shall employ two canonical transformations: in the first one, $\Lambda$ is replaced by the mass $M$ of the dust cloud; in the second one, $M$ is replaced by the Schwarzschild Killing time $T$. After the implementation of the first canonical transformation, a boundary term similar to the one in \cite{Kuchar:1994zk} arises, with the difference that here the boundary term will not vanish at $r_s$ according to the Kucha\v{r} fall-off conditions.

Let us consider the first transformation $(\Lambda,R,P_{\Lambda},P_R)\rightarrow (M,\vb{R},P_M,\vb{P}_R)$ with 
\begin{align}
\label{canonic traf1}
    M&=\frac{R}{2}\left(1-F\right),\\
    \label{canonic trafo2}
    R&=\vb{R},\\
    \label{canonic trafo3}
    P_M&=\frac{\Lambda P_{\Lambda}}{RF},\\ 
    \vb{P}_R&=P_R-\frac{\Lambda P_{\Lambda}}{2R}-\frac{\Lambda P_{\Lambda}}{2RF}\\ \nonumber&-\frac{(\Lambda P_{\Lambda})'RR'-(RR')'\Lambda P_{\Lambda}}{R\Lambda^2 F},
    \label{canonic trafo 4}
\end{align}
where 
\begin{equation}
    F:=\frac{R'^2}{\Lambda^2}-\frac{P_{\Lambda}^2}{R^2}.
\end{equation}
The difference in the Liouville forms after the canonical transformation is
\begin{equation}
\begin{aligned}
    \int_{r_s}^{\infty} dr\left( P_R \dot{R}+ P_{\Lambda}\Dot{\Lambda}\right)-&\int_{r_s}^{\infty} dr \left(\vb{P}_R\Dot{R}+P_M\Dot{M}\right)\\
    &=R\Dot{R} \ln{\abs{\frac{RR'-\Lambda P_{\Lambda}}{RR'+\Lambda P_{\Lambda}}}}.
    \end{aligned}
    \label{boundary term chap 4}
\end{equation}
Evaluating the boundary term using \eqref{match 4.1} and \eqref{match 4.2} gives
\begin{equation}
    \mathrm{boundary\ term}=\frac{a\Dot{a}r_s^2}{2}\ln{\abs{\frac{a-\frac{r_s P}{3V_s\sqrt{1-kr_s^2}}}{a+\frac{r_s P}{3V_s\sqrt{1-kr_s^2}}}}}.
    \label{boundary chapter 4 termi}
\end{equation}
This boundary term \eqref{boundary chapter 4 termi} will not disappear for $r\rightarrow r_s$, which is problematic since it prevents the action from being canonical. Getting rid of this term would automatically provide different time coordinates for different interior curvatures, without any external input in the canonical formulation. Furthermore, it would help us to find a deparametrized Hamiltonian for the OS-model (from the point of view of the stationary observer). This will be discussed shortly.

The dust cloud mass behaves as 
\begin{equation}
    M(r_s)=\frac{r_s^3}{3V_s}\left(\frac{P^2}{6V_sa}+\frac{3V_s}{2}ka\right)=\frac{r_s^3}{3V_s}\left(\Bar{P}_{\tau}-\Bar{H}\right).
    \label{dust cloud mass 4.}
\end{equation}
The next step is to implement a second canonical transformation, $(M,P_M)\rightarrow(T,P_T)$ . We identify $P_M=T'$, hence 
\begin{equation}
    T=T_++\int_r^{\infty} dr' P_M\,.
\end{equation}
Let us consider the relevant Liouville term, 
\begin{equation}
    -M_+\Dot{T}_+ + \int_{r_s}^{\infty} P_M \Dot{M}=-M_+\Dot{T}_-\int_{r_s}^{\infty} dr\, T'\Dot{M}.
    \label{canonical chapteri32}
\end{equation}
The right-hand side of \eqref{canonical chapteri32} can be rewritten as
\begin{equation}
\begin{aligned}
   & -M\Dot{T}-\int dr\, M' \Dot{T}\\&+\dv{}{t}\left(MT-M_+T_++\int_{r_s}^{\infty} dr\, M' T\right).
   \label{ronaldinho}
   \end{aligned}
\end{equation}
Ignoring the total time derivative in \eqref{ronaldinho} and identifying $P_T=-M'$, the boundary term further simplifies to 
\begin{equation}
    -M\Dot{T}-\int_{r_s}^{\infty}P_T \,\Dot{T} dr.
    \label{ronaldinho 2}
\end{equation}
Kucha\v{r} showed in \cite{Kuchar:1994zk} that one can replace the complicated system of constraints $H=H_r=0$ with the considerably simpler system $P_T=\vb{P}_R=0$; hence the full action after all these simplifications reads 
\begin{equation}
    \begin{aligned}
     \int dt &\left(P\Dot{a}+\Bar{P}_{\tau}\Dot{\tau}+\frac{a\Dot{a}r_s^2}{2}\ln{\abs{\frac{a-\frac{r_s P}{3V_s\sqrt{1-kr_s^2}}}{a+\frac{r_s P}{3V_s\sqrt{1-kr_s^2}}}}}-\Bar{M}\Dot{T}-\Bar{N}\Bar{H}\right)\\
     &+\int dt \int_{r_s}^{\infty} dr \left(\vb{P}_R\Dot{R}+P_T\Dot{T}-N^R\vb{P}_R-N^TP_T\right).
    \end{aligned}
    \label{real final 4.0 action}
\end{equation}

It is obvious that this action is not canonical because of the boundary terms which arise from equations \eqref{ronaldinho 2} and \eqref{boundary chapter 4 termi}. To make this action canonical, we choose $\Bar{T}$ in such a way that additional boundary terms in the Liouville form disappear. In \cite{tim1}, this problematic boundary term was for the flat case $(k=0)$ re-written as a total time derivative; unfortunately, this is not possible for the curved cases $(k=\pm 1)$. In the next subsection we shall present a new method to resolve this issue.

For the action to become canonical, the additional boundary terms need to vanish. We therefore enforce 
\begin{equation}
    \begin{aligned}
     \frac{a\Dot{a}r_s^2}{2}\ln{\abs{\frac{a-\frac{r_s P}{3V_s\sqrt{1-kr_s^2}}}{a+\frac{r_s P}{3V_s\sqrt{1-kr_s^2}}}}}-\Bar{M}\Dot{T}=0.
    \end{aligned}
    \label{boundary stuff}
\end{equation}
In \cite{tim1} it was shown for the flat case that Painlev\'e-Gullstrand (PG) coordinates appear if one wants to find a suitable $\Bar{T}$ in order to render the action canonical. There it was possible to directly calculate $\Bar{T}$ and rewrite it using PG-coordinates. We shall follow here a different route along which one can generalize this method to include the open and closed cases as well. Though, this approach comes with a price: we are not able to directly calculate $\Bar{T}$. Instead, we will find its derivative to be derivatives of 
the PG-, the generalized PG-\cite{Martel:2000rn}, and the Gautreau-Hoffman \cite{gautreau} coordinates, depending on the choice of $k$. These coordinates are therefore suitable candidates for $\Bar{T}$, but we could perform a constant shift to them as well, since we are only dealing with derivatives. 

We now first rescale our variables to $R=ar_s$ and $P_R=p/r_s$ and rewrite \eqref{boundary stuff} to find
\begin{equation}
    \frac{R\Dot{R}}{2}\ln{\abs{\frac{\Bar{R}\sqrt{1-kr_s^2}-\frac{r_S^3}{3V_s}\Bar{P}_R}{\Bar{R}\sqrt{1-kr_s^2}+\frac{r_s^3}{3V_s}\Bar{P}_R}}}-M\dot{\Bar{T}}=0.
    \label{boundary 4.2.2}
\end{equation}
Using \eqref{dust cloud mass 4.}, we moreover get
\begin{equation}
\begin{aligned}
    &\frac{R\Dot{R}}{2}\ln{\abs{\frac{\Bar{R}\sqrt{1-kr_s^2}-\frac{r_S^3}{3V_s}\Bar{P}_R}{\Bar{R}\sqrt{1-kr_s^2}+\frac{r_s^3}{3V_s}\Bar{P}_R}}}\\&-\left(\frac{r_s^6}{18v_s^2}\frac{\bar{P_R}^2}{\bar{R}}+\frac{kr_s^2}{2}\Bar{R}\right)\dot{\Bar{T}}=0.
    \label{boundary 4.2.3}
    \end{aligned}
\end{equation}
To get $\Bar{T}$ from \eqref{boundary 4.2.3} it should assume the form
\begin{equation}
    \begin{aligned}
        &A(\Bar{R},\Bar{P}_R)\Dot{B}(\Bar{R},\Bar{P}_R,\Bar{T})+\Dot{C}(\Bar{R},\Bar{P}_R,\Bar{T})=\\ &A \left(\pdv{B}{\Bar{R}}+\pdv{C}{R}\right)\dot{\Bar{R}}+\left(A\pdv{B}{\Bar{P}_R}+ \pdv{C}{\Bar{P}_R}\right)\Dot{\Bar{P}}_R&\\&+ \left(A\pdv{B}{\Bar{T}}+\pdv{C}{\Bar{T}}\right)\Dot{\Bar{T}}.
    \end{aligned}
    \label{general form of T}
\end{equation}
Comparison of \eqref{boundary 4.2.3} and \eqref{general form of T} leads to the following set of differential equations:
\begin{align}
    A\pdv{B}{\Bar{R}}+\pdv{C}{R}=&\frac{R}{2}\ln{\abs{\frac{\Bar{R}\sqrt{1-kr_s^2}-\frac{r_S^3}{3V_s}\Bar{P}_R}{\Bar{R}\sqrt{1-kr_s^2}+\frac{r_s^3}{3V_s}\Bar{P}_R}}},
    \label{1 diff34}\\A\pdv{B}{\Bar{P}_R}+\pdv{C}{\Bar{P}_R}=&0, \label{2 diff35}\\A\pdv{B}{\Bar{T}}+\pdv{C}{\Bar{P}_R}=& -\frac{r_s^6}{18V_s^2}\frac{\Bar{P}_R^2}{\Bar{R}}-\frac{kr_s^2}{2}\Bar{R}
.\label{3.diff 36}
\end{align}
We first differentiate \eqref{1 diff34} with respect to $\Bar{T}$ and \eqref{3.diff 36} with respect to $\Bar{R}$. Combining the results yields
\begin{equation}
\pdv{A}{\Bar{R}}\pdv{B}{\Bar{T}}=\frac{r_s^6}{18V_s^2} \frac{\Bar{P}_R^2}{\Bar{R}^2}-\frac{kr_s^2}{2}. \label{4 diff 37}
\end{equation}
Similarly, we combine equations \eqref{2 diff35} and \eqref{3.diff 36} after differentiating them with respect to $\Bar{T},\Bar{P}_R$, respectively, and obtain
\begin{equation}
\pdv{A}{\Bar{P}_R}\pdv{B}{\Bar{T}}=-\frac{r_s^6}{9V_s^2}\frac{\Bar{P}_R}{\Bar{R}}.
\label{5.diff 38}
\end{equation}
By dividing equations \eqref{4 diff 37}, \eqref{5.diff 38}, we find
\begin{equation}
    \pdv{A}{\Bar{R}}=\left(-\frac{\Bar{P}_R}{2R}+k\frac{9V_s^2}{2r_s^4} \frac{\Bar{R}}{\Bar{P}_R}\right)\pdv{A}{\Bar{P}_R}.
    \label{39}
\end{equation}
This equation can be used to obtain  
\begin{equation}
    A(\Bar{R},\Bar{P}_R)=c_1\left(\frac{\Bar{P}_R^2}{\Bar{R}}+k \frac{9V_s^2}{r_s^4}\Bar{R}\right)+c_2,
    \label{diff 40}
\end{equation}
where $c_1,c_2$ are constants of integration. We can now insert \eqref{diff 40} into \eqref{5.diff 38} and thereby find for $B(\Bar{R},\Bar{P}_R,\Bar{T})$ the expression
\begin{equation}
   B(\Bar{R},\Bar{P}_R,\Bar{T})=-\frac{r_s^6}{18c_1V_s^2} \Bar{T}+D(\Bar{R},\Bar{P}_R).
   \label{41}
\end{equation}
We note that $D(\Bar{R},\Bar{P}_R)$ has emerged from an integration with respect to $\Bar{T}$. Since $A(\Bar{R},\Bar{P}_R)$ and $B(\Bar{R},\Bar{P}_R,\Bar{T})$ are now known, we can use \eqref{3.diff 36} to find 
\begin{equation}
    C(\Bar{R},\Bar{P}_R,\Bar{T})=\frac{c_2r_s^6}{18c_1V_s^2}\Bar{T}+E(\Bar{R},\Bar{P}_R).
    \label{42}
\end{equation}
By using \eqref{diff 40}, \eqref{41}, and \eqref{42}, we can reformulate \eqref{general form of T} as 
\begin{equation}
    \begin{aligned}
        &A(\Bar{P}_R,\Bar{R})\Dot{B}(\Bar{R},\Bar{P}_R,\Bar{T})+\Dot{C}(\Bar{R},\Bar{P}_R,\Bar{T})
    \\& \left(\frac{\Bar{P}_R}{\Bar{R}}+k \frac{9V_s^2}{r_s^4}\Bar{R}\right)\left(-\frac{r_s^6}{18V_s^2}\dot{\Bar{T}}+c_1\Dot{D}(\Bar{R},\Bar{P}_R)\right)\\& +\left(c_2 \Dot{D}(\Bar{R},\Bar{P}_R)+\Dot{E}(\Bar{R},\Bar{P}_R)\right).
    \end{aligned}
\end{equation}
It is possible to absorb $c_1$ into $\Dot{D}(\Bar{R},\Bar{P}_R)$. In other words, we can set $c_1=1$ without loss of generality. Likewise, it is possible to set $c_2=0$ because we can absorb $c_2\Dot{D}(\Bar{R},\Bar{P}_R)$ into $\Dot{E}(\Bar{R},\Bar{P}_R)$. Going forward we can get new information by differentiating \eqref{1 diff34} with respect to $\Bar{P}_R$ and combining it with the derivative with respect to $\Bar{R}$ of equation \eqref{2 diff35}. This will lead to a new equation for $B(\Bar{R},\Bar{P}_R,T)$, which can equivalently be formulated in terms of  $D(\Bar{R},\Bar{P}_R)$, since $\pdv{B}{\Bar{R}}=\pdv{D}{\Bar{R}}$ and $\pdv{B}{\Bar{P}_R}=\pdv{D}{\Bar{P}_R}$. With this, we arrive at 
\begin{equation}
    \begin{aligned}
        &\frac{2 \Bar{P}_R}{R}\pdv{D}{\Bar{R}}- \left(-\frac{\Bar{P}_R^2}{\Bar{R}^2}+k\frac{9V_s^2}{r_s^4}\right)\pdv{D}{\Bar{P}_R}\\&= \frac{\frac{r_s^3}{3V_s}\sqrt{1-kr_s^2}\Bar{R}^2}{\frac{r_s^6}{9V_s^2}\Bar{P}_R^2-\left(1-kr_s^2\right)\Bar{R}^2}.
    \end{aligned}
    \label{poisson 1}
\end{equation}
We can now rewrite the Hamiltonian for the comoving observer and the mass of the dust cloud in terms of the rescaled variables $R=ar_s$, $\Bar{P}_R=\frac{p}{r_s}$ to find 
\begin{align}
 \Bar{H}=&-\frac{r_s^3}{6V_s}\frac{\Bar{P}_R^2}{R}-k\frac{3V_s}{2r_s}\Bar{R}+\Bar{P}_{\tau},\label{rescaled hamiltonian}\\
 \Bar{M}=&\frac{r_s^6}{18V_s^2} \frac{\Bar{P}_R^2}{\Bar{R}}+\frac{kr_s^2}{2}\Bar{R}.\label{rescaled mass1}
 \end{align}
 The redefined mass \eqref{rescaled mass1} can be used to obtain a useful expression for the momentum $\Bar{P}_R$,
 \begin{equation}
     \Bar{P}_R^2=\frac{9V_s^2}{r_s^6}\left(\frac{2\Bar{M}}{\Bar{R}}-kr_s^2\right)\Bar{R}^2.
     \label{useful relation for PR}
 \end{equation}
 Calculating the derivative of the new Hamiltonian \eqref{rescaled hamiltonian}, we find
 \begin{align}
     -\frac{6V_s}{r_s^3}\pdv{\Bar{H}}{\Bar{P}_R}=&\frac{2\Bar{P}_R}{\Bar{R}},\label{derivative1}\\
     -\frac{6V_s}{r_s^3}\pdv{\Bar{H}}{\Bar{R}}=&-\frac{\Bar{P}_R^2}{\Bar{R^2}}+k\frac{9V_s^2}{r_s^4}.\label{derivative 2}
 \end{align}
 This allow us to identify a Poisson bracket on the left-hand side of \eqref{poisson 1}, which can be rewritten as 
 \begin{equation}
     \begin{aligned}
         -\frac{6V_s}{r_s^3}\qty\bigg{D,\Bar{H}}=&-\frac{6V_s}{r_s^3}\pdv{D}{\tau}\\ &=\frac{\frac{r_s^3}{3V_s}\sqrt{1-kr_s^2}\Bar{R}^2}{\frac{r_s^6}{9V_s^2}\Bar{P}_R^2-\left(1-kr_s^2\right)\Bar{R}^2}\,.
     \end{aligned}
     \label{poisson 2}
 \end{equation}
It should be noted that replacing the Poisson bracket with the derivative of $D$ with respect to $\tau$ is 
only valid after going on-shell. Furthermore, we can identify $\tau$ as the observer's proper time. This will be further substantiated once we identify the different coordinates for the different interior curvatures. One caveat is that the derivatives in this equation are with respect to label time. In contrast, in equation \eqref{poisson 2} the time parameter $\tau$ appears because we make use of a Poisson bracket for its formulation. We can take this into account by introducing 
$\Dot{\tau}$; hence, 
\begin{equation}
\begin{aligned}
    \frac{\Bar{\Dot{T}}}{\Bar{\Dot{\tau}}}=\frac{18V_s^2}{r_s^6}\pdv{\Bar{T}}{\tau}=\frac{\frac{r_s^3}{3V_s}\sqrt{1-kr_s^2}\Bar{R}^2}{\frac{r_s^6}{9V_s^2}\Bar{P}_R^2-\left(1-kr_s^2\right)\Bar{R}^2}.
    \end{aligned}
    \label{tau derivative}
\end{equation}

%%%%%%%%%%%%%%%%%%%%%%%%%%%%%%%%%%%%%%%%%%%%%%%

\section{Coordinate systems}
\label{chapter 4}
\subsection{Painlev\'e-Gullstrand coordinates}
The PG-coordinates already mentioned above yield a coordinate system which is regular at the event horizon and singular at $R=0$. We briefly introduce them here and refer to e.g. \cite{Martel:2000rn} for more details. 

We start from the four-velocity $u^{\alpha}=(\Dot{t},\Dot{R},0,0)$ and its covariant components $u_{\alpha}=\left(-1,-\frac{\sqrt{1-F}}{F},0,0\right)$. This means that $u_{\alpha}$ is equal to the gradient of some arbitrary function $T$,
\begin{equation}
    u_{\alpha}=-\pderivative{\alpha}T,
\end{equation}
where 
\begin{equation}
    T=t+\int\frac{\sqrt{1-F}}{F} dr.
    \label{ap,8}
\end{equation}
Differentiating \eqref{ap,8} gives, with $F=1-\frac{2M}{R}$,
\begin{equation}
dt=dT-\frac{\sqrt{\frac{2M}{R}}}{1-\frac{2M}{R}} dr.
\label{ap,9}
\end{equation}
Inserting \eqref{ap,9} into the Schwarzschild line element
\begin{equation}
    dS^2=-\left(1-\frac{2M}{R}\right)dt^2+\left(1-\frac{2M}{R}\right)dR^2+R^2 d\Omega^2,
    \label{app.10}
\end{equation}
one finds by some straightforward manipulations the line element
\begin{equation}
    dS^2=-dT^2+\left(dR+\sqrt{\frac{2M}{R}}dT\right)^2+R^2 d\Omega^2.
    \label{ap.11}
\end{equation}
This metric \eqref{ap.11} is non-diagonal and regular at $R=2M$, but singular at 
$R=0$. Its disadvantage compared to Kruskal-Szekeres coordinates is that it only covers two regions of the extended Schwarzschild space-time, but the interesting property of \eqref{ap.11} is that the surface of constant $T$ is intrinsically flat: $dT=0$ means $dS^2=dR^2+R^2 d\Omega^2$, which is the metric of the flat three-dimensional space in spherical coordinates. Moreover, it corresponds to an observer who starts with zero initial velocity at infinity; for more details, see \cite{Poisson:2009pwt}. 

\subsection{Generalized PG-coordinates}
\label{generalized pg}
In order to derive generalized PG-coordinates we follow the same procedure as in the last subsection. We find that $u_{\alpha}=-\frac{1}{\sqrt{p}}\pderivative{\alpha}T$, where $T$ is an arbitrary function that reads 
\begin{equation}
    T=t+\int \frac{\sqrt{1-PF}}{F} dR\,.
    \label{Gpg}
\end{equation}
Inserting 
\begin{equation}
    dt=dT-F^{-1}\sqrt{1-PF}\,dR
\end{equation}
into \eqref{app.10} yields
\begin{equation}
    dS^2=-\frac{1}{p}dT^2+P\left(dR+\frac{1}{P}\sqrt{1-PF}dT\right)^2+R^2 d\Omega^2.
\end{equation}
Note that when $T=\mathrm{constant}$, $dT$ will disappear and we end up with a curved 
three-dimensional space which is no longer flat,
\begin{equation}
    dS^2=pdR^2+R^2 d\Omega^2.
\end{equation}
The coefficient $P<1$ is responsible for the appearance of curvature. The only non-vanishing component of the Riemann tensor is 
\begin{equation}
    R^{\phi}_{\theta \phi \theta}=-\frac{1-p}{p}.
\end{equation}
The limit $p\rightarrow 1$ reproduces the flat metric. We find, moreover, that observers with initial non-vanishing velocity correspond to a curved three-metric and we thus cover here the open case, $k=-1$; for more details, see \cite{Martel:2000rn}. 

\subsection{Gautreau-Hoffman coordinates}
The closed $(k=1)$ case which corresponds to an initially vanishing velocity at finite radius $r_s$ is covered by another set of coordinates called Gautreau-Hoffman coordinates,
\begin{equation}
    dS^2=-dT^2+F^{-1}\left(dR-\sqrt{\frac{2M}{R}-\frac{2M}{R_i}}\,dT\right)^2,
\end{equation}
with $T$ given by
\begin{equation}
    t=FT-\int_{R_i}^{R} dy \frac{\sqrt{\frac{2M}{y}-\frac{2M}{R_i}}}{1-\frac{2M}{y}}.
    \label{GH}
\end{equation}
These coordinates are very similar to the coordinates of the last subsection \ref{generalized pg}. They constitute a one-parameter family of coordinate systems, and their parameter $R_i$ is related to $P$ by \cite{Martel:2000rn} 
\begin{equation}
    R_i=\frac{2MP}{P-1}.
\end{equation}
The point is that formally the coordinates of section \ref{generalized pg} are identical to 
Gautreau-Hoffman coordinates for $R_i<0$. For $R_i>0$, which is the case studied by Gautreau and Hoffman, the surfaces of constant time extend only up to $r=R_i$, that is, they do not reach infinity. We will not discuss this here, but refer to \cite{Martel:2000rn} and \cite{gautreau} for more information. 

\section{General time coordinate}
\label{chapter 5}
Comparing \eqref{ap,8}, \eqref{Gpg}, and \eqref{GH}, one can generalize these three equations to
\begin{equation}
    \Bar{T}=\sqrt{p_k}\tau+\int dR \frac{\sqrt{1-p_kF}}{F},
    \label{general time1}
\end{equation}
with 
\begin{align}
\label{ap.24}
    p_k&=\frac{1}{1-kr_s^2},\\
    F&=1-\frac{2\Bar{M}}{R}.
\end{align}
Furthermore, we have
\begin{equation}
    P=p_k\equiv \frac{1}{1-kr_s^2}\,.
\end{equation}
where $k$ stands again for the curvature values $\pm1,0$. We emphasize that \eqref{general time1} covers open, closed, and flat spatial sections. It will play a central role in our analysis to develop a Hamiltonian for the static OS-model. Let us now calculate the derivative of the Killing time \eqref{general time1} with respect to proper time $\tau$ to find
\begin{equation}
    \pdv{\Bar{T}}{\tau}=\sqrt{p_k}+\epsilon \pdv{\Bar{R}}{\tau} \frac{\sqrt{1-p_k F}}{F},
    \label{T derivative 1}
\end{equation}
where $F=1-\frac{2M}{R}$. Using the definition of mass $\Bar{M}$, equation \eqref{rescaled mass1}, we can rewrite equation \eqref{T derivative 1} such that it becomes
\begin{equation}
\begin{aligned}
    \pdv{\Bar{T}}{\tau}=\frac{1}{\sqrt{1-kr_s^2}}\left( 1-\frac{r_s^6}{9V_s^2}\frac{\Bar{P}_R^2}{2\Bar{M}\Bar{R}-\Bar{R}^2}\right).
\end{aligned}
\end{equation}
Employing \eqref{useful relation for PR} we arrive at
\begin{equation}
    \begin{aligned}
        \pdv{\Bar{T}}{\tau}=&\frac{1}{\sqrt{1-kr_s^2}}\left(1-\frac{2\Bar{M}\Bar{R}-kr_s^2\Bar{R}^2}{2\Bar{M}\Bar{R}-\Bar{R}^2}\right)\\=&-\frac{\sqrt{1-kr_s^2}\Bar{R}^2}{\frac{r_s^6}{9V_s^2}\Bar{P}_R^2-\left(1-kr_s^2\right)\Bar{R}^2},
    \end{aligned}
    \label{T derivative finale}
\end{equation}
which is identical to equation \eqref{tau derivative}. Therefore, we can conclude that the three coordinate systems are indeed suitable choices for making the action \eqref{real final 4.0 action} truly canonical again. This is an important result. The procedure of eliminating the boundary term \eqref{boundary stuff} also provides us with the insight that we get different time coordinates for different interior curvatures, without any external input in the canonical formulation. This is indeed very interesting, since it seems that the canonical formulation intrinsically knows about these coordinate time transformations. Furthermore, it will help us to find a deparametrized Hamiltonian for the OS-model (from the point of view of the stationary observer). This will be discussed in section \ref{static observer hamiltonian 6}.

\section{Transformation of time coordinates and derivation of the Hamiltonian for a stationary observer}
\label{static observer hamiltonian 6}
The canonical transformation between proper time $\tau$ and Schwarzschild time $\Bar{T}$ can be found by using a generating function of the third kind, $\Omega(\Bar{P}_R,\Bar{P}_{\tau},\Bar{R},\Bar{T})$, which allows us to transform the coordinates, momenta, and the Hamiltonian such that it is appropriate for a stationary observer; for more details on generating functions, see e.g. \cite{goldstein2002classical}. We have
\begin{align}
\label{generating1}
    \Bar{R}&=-\pdv{\Omega(\Bar{P}_R,\Bar{P}_{\tau},\Bar{\vb{R}},\Bar{T})}{\Bar{P}_{\Bar{R}}},\\
    \tau&=-\pdv{\Omega(\Bar{P}_R,\Bar{P}_{\tau},\Bar{\vb{R}},\Bar{T})}{\Bar{P}_{\tau}}.
    \label{generating 2}
\end{align}
Integrating these two equations, we find 
\begin{equation}
\begin{aligned}
    \Omega(\Bar{P}_{\Bar{R}}, \Bar{P}_{\tau},\Bar{R},\Bar{T})=&-\Bar{P}_{\Bar{R}}R-\frac{\Bar{P}_{\tau}\Bar{T}}{\sqrt{p_k}}\\&+\frac{\epsilon}{\sqrt{p_k}}\int d\Bar{P}_{\tau}\int dR\, \frac{\sqrt{1-p_kF}}{F}\\&+\Theta(R,\Bar{T});
    \end{aligned}
    \label{generating stuff for summary}
\end{equation}
the values $\epsilon=\pm1$ indicate dust collapse and expansion, respectively; $\Theta$ was introduced in order to implement the property that $\Omega$ is only defined by derivatives with respect to $\Bar{P}_{\Bar{R}}, \Bar{P}_{\tau}$. 

We shall now calculate the momenta conjugate to $\Bar{R},\Bar{T}$ from the generating function,
\begin{widetext}
\begin{align}
\Bar{P}_{\Bar{R}}&=-\pdv{\Omega}{\Bar{R}}\label{stationary P_R}\\ \nonumber 
&=\Bar{P}_{\Bar{R}}+\frac{3\epsilon V_s}{\sqrt{p_k}r_s^3}\Bar{R}\sqrt{1-p_k\left(1-\frac{\frac{2r_s^3}{3V_s}\Bar{P}_{\tau}}{\Bar{R}}\right)}-\pdv{\Theta}{\Bar{R}}\\- \,\nonumber &\!\frac{3\epsilon V_s \Bar{R}}{\sqrt{p_k}r_s^3}
\begin{cases}
\tanh^{-1}{\left(\sqrt{1-p_k\left(1-\frac{2\frac{r_s^3}{3V_s}\Bar{P}_{\tau}}{\Bar{R}}\right)}\right)},\!&\Bar{R}>\frac{2r_s^3}{3V_s}\Bar{P}_{\tau}\\
\coth^{-1}{\left(\sqrt{1-p_k\left(1-\frac{2\frac{r_s^3}{3V_s}\Bar{P}_{\tau}}{\Bar{R}}\right)}\right)},\!&\Bar{R}<\frac{2r_s^3}{3V_s}\Bar{P}_{\tau}
\end{cases}.
\end{align}
\end{widetext}
 A similar calculation for $\Bar{P}_{\Bar{T}}$ leads to
\begin{equation}
    \Bar{P}_{\Bar{T}}=\pdv{\Omega}{\Bar{T}}=\frac{\Bar{P}_{\tau}}{\sqrt{p_k}}-\pdv{\Omega}{\Bar{T}}\,.
    \label{76}
\end{equation}
In order to preserve the intuitive relationship between the momentum and dust cloud mass, we can choose $\Theta=0$, which implies 
$\Bar{P}_{\Bar{T}}=\frac{\Bar{P}_{\tau}}{\sqrt{p_k}}$, leading to $\Bar{P}_{\Bar{T}}=\Bar{P}_{\tau}$ for the flat case. Inserting \eqref{rescaled hamiltonian} into \eqref{stationary P_R} and rewriting the resulting equation, we find the Hamiltonian for the stationary observer,
\begin{equation}
\begin{aligned}
    &\Bar{H}_{\Bar{T}}= \Bar{P}_{\Bar{T}}+\\ & \frac{3\Bar{R}\,V_s}{2\sqrt{p_k}r_s^3 p_k}  \begin{cases}
    1-p_k-\tanh^2{\left(\frac{r_s^3\Bar{P}_{\Bar{R}}}{3V_s\Bar{R} }\sqrt{p_k}\right)},& \Bar{R}>\frac{2 r_s^3}{3V_s}\Bar{P}_{\tau}\\
     1-p_k-\coth^2{\left(\frac{r_s^3\Bar{P}_{\Bar{R}}}{3V_s\Bar{R} }\sqrt{p_k}\right)},& \Bar{R}<\frac{2 r_s^3}{3V_s}\Bar{P}_{\tau}
    \end{cases}.
    \label{stationary hamiltonian}
\end{aligned}
\end{equation}
The split between collapse and expansion has disappeared because $\epsilon$ does not appear in the above Hamiltonian. The split at the horizon $\Bar{R}=\frac{2r_s^3}{3V_s \sqrt{p_k}}\Bar{P}_{\Bar{T}}$, however, still exists. Furthermore, this Hamiltonian has a branching point at $P_R\rightarrow \pm\infty$; concerning the behavior of trajectories , the Hamilton equations of motion describe radial Schwarzschild geodesics in Killing time. Therefore they show the usual asymptotic approach towards the horizon, and there is no transition between the two branches at finite Killing time. Furthermore, we recall that for the flat case we have $p_k=1,V_s=\frac{r_s^3}{3}$, which will exactly reproduce the stationary Hamiltonian derived for flat space in \cite{tim1}. 

Now that we have derived the Hamiltonian for static observers in the classical OS-model (non-marginal case), we can discuss its quantization. This will be done in the next section. We would like to end this section by discussing how we can deparametrize \eqref{stationary hamiltonian} with respect to Killing time $T$. In its current form \eqref{stationary hamiltonian}, the effective Hamiltonian still depends on $P_T$ through the position of split inside and outside of the horizon. Having the explicit split in the constraints is unnecessary, since the conditions $R<\frac{2r_s^3}{3V_s}$ and $R>\frac{2r_s^3}{3V_s}$ are automatically implemented by the constraints themselves. We can then identify an effective Hamiltonian 
\begin{equation}
\begin{aligned}
{H}=\frac{3\Bar{R}\,V_s}{2\sqrt{p_k}r_s^3 p_k} \, \begin{cases}
    1-p_k-\tanh^2{\left(\frac{r_s^3\Bar{P}_{\Bar{R}}}{3V_s\Bar{R} }\sqrt{p_k}\right)}\\
     1-p_k-\coth^2{\left(\frac{r_s^3\Bar{P}_{\Bar{R}}}{3V_s\Bar{R} }\sqrt{p_k}\right)}
    \end{cases},
    \end{aligned}
    \label{stationary hamiltonian finale 4}
    \end{equation}
which is a multivalued one. Due to the unusual form of this Hamiltonian \eqref{stationary hamiltonian finale 4}, we cannot simply proceed with the standard Dirac canonical quantization program. Therefore, in the next section, we shall discuss an alternative scheme for quantization. We note that the procedure employed in this section heavily takes advantage of the simplicity of the OS-model and cannot be easily adopted to, for example, the LTB collapse model. Switching the observer as above applies to its outermost dust shell, but it is unclear to us how this can be generalized to the other shells. For details concerning the LTB-model, see \cite{LTB}.
%%%%%%%%%%%%%%%%%%%%%%%%%%%%%%%%%%%%%%%%%%%%%%%%%%%%%%%%%
\section{Quantization}
\label{chapter 7}
In the last section \ref{static observer hamiltonian 6} we have derived a Hamiltonian for a static observer who is located outside the black hole. Due to the unusual form of the Hamiltonian the effort to quantize \eqref{stationary hamiltonian finale 4} employing the Dirac quantization scheme seems impossible. This motivated the authors in \cite{tim} to apply another quantization scheme known as Affine Coherent State Quantization (ACSQ). The procedure is as follows. We first identify the phase space by a Lie group. Then we consider a unitary irreducible representation of it on a Hilbert space and let it act on a fixed state. This leads to a family of \emph{coherent states}, which provide us with a resolution of the identity. Phase space functions can thus be mapped to operators by inserting them into this resolution. This procedure allows for more complicated phase space functions to be paired up with operators. The most desirable feature of this scheme is that if the phase space functions are at least semibounded, the resulting operator after quantization is self-adjoint \cite{Gazeau}. In this section, we shall give a brief introduction to ACSQ mainly following \cite{almeida2018three}; for more details, see \cite{Gazeau:2015nkc}, \cite{Gazeau:2014qqa}, and \cite{klaudera}. This method has already been applied to quantum cosmology where it was shown that the classical singularity can be replaced by a bounce \cite{Bergeron:2013ika}, \cite{Gozdz:2019aoa}.

\subsection{The affine group and its representation}
\label{affine group representation}
The half plane can be viewed as the phase space of a particle moving on the half line. Let the upper half plane $\Pi_+=\qty{(R,P)|R>0,P\in \mathbb{R}}$ be equipped with the measure $dPdR$ with multiplication
\begin{equation}
    (R,P)(R_0,P_0)=\left(RR_0,\frac{P_0}{R}+P\right).
\end{equation}
The unity is $(1,0)$ and the inverse is 
\begin{equation}
    (R,P)^{-1}=(\frac{1}{R},-RP).
\end{equation}
The upper half plane $\Pi_+$ can be regarded as the affine group $\textit{aff}_+(\mathbb{R})$ of the real line,
\begin{equation}
    \mathbb{R}\ni y \rightarrow (P,R).y=\frac{y}{R}+P.
\end{equation}
We choose, moreover, the standard Liouville phase space measure $dPdR$ because it is invariant with respect to the left action of the affine group on itself,
\begin{align}
    \textit{aff}_+(\mathbb{R})\ni (R,P)\rightarrow (R_0,P_0)(R,P)&=(R',P'),\\
    dR'dP'&=dRdP.
\end{align}
The affine group has two non-equivalent unitary irreducible representations $U_{\pm}$ \cite{almeida2018three}. Without loss of generality we choose $ U\equiv U_+$,
\begin{equation}
    U(P,R)\psi(x)=\frac{e^{\frac{i}{\hbar}px}}{\sqrt{R}}\psi\left(\frac{x}{R}\right).
\label{representation}
\end{equation}
Equation \eqref{representation} means that this group is acting on the Hilbert space $\mathcal{H}=L^2(R_+,dx)$ by using the above representation.

\subsection{Affine covariant integral quantization}
\label{affine covariant integral quantization}
Let $G$ be a compact Lie group with left Haar measure $d\mu(g),\, d\mu(g_0g)=d\mu(g)$ for all $g_0\in G$, $g\rightarrow U(g)$ being a unitary irreducible representation (UIR) of $G$ in Hilbert space. Let us consider a bounded operator $M$ on $\mathcal{H}$, 
\begin{equation}
    R=\int M(g) d\mu(g).
\end{equation}
From left invariance we have 
\begin{equation}
    U(g)RU^{\dagger}(g)=\int_G M(g_0g) d\mu(g)=R.
\end{equation}
We then employ Schur's lemma which states: Let $G$ be a group with its unitary irreducible representation on a vector space V, if $U(g)M(g)U^{\dagger}(g)=M(g)$ for all $g\in G$, then $M$ is a multiple of identity. Using this lemma we get 
\begin{align}
   \int M(g) d\mu(g)&=c_M I,\\
   \label{covariant i q}
   \int M(g) d\nu(g)&=I,\\
   \frac{d\mu(g)}{c_M}&=d\nu(g).
\end{align}
Moreover, the subsequent quantization of complex valued functions on $G$ reads 
\begin{equation}
\label{ciq 1}
    f\rightarrow A_f=\int_G M(g) f(g) d\nu(g).
\end{equation}
The linear map of a $\textit{function}\rightarrow \textit{operator in}\ \mathcal{H}$ is covariant in the sense that 
\begin{equation}
    U(g)A_f U^{\dagger}(g)=A_{U(g)f}
\end{equation}
A semiclassical analysis of the operator $A_f$ can be implemented through the study of a new function called lower symbol of $f$, denoted by $\Check{f}$. These are the generalization of the covariant symbol introduced by Berezin \cite{brazil}. Suppose that $M$ is a non-negative unit-trace operator $M=\rho$ on $\mathcal{H}$. Then the operators $\rho(g)$ are also densities, and this allows to write the function $\Check{f}(g)$ as 
\begin{equation}
    \Check{f}(g)\equiv \Check{A}_f=\int_G \Tr{\rho(g)\rho(g')}f(g')d\nu(g').
    \label{ciq 2}
\end{equation}
\subsection{Affine coherent state quantization}
\label{acsq section}
Let us implement the affine covariant integral quantization, which is described in \ref{affine covariant integral quantization} in its generality, by restricting the method to rank one density operators $\rho= \ket{\phi}\bra{\phi}$, where the $\phi$ are unit norm states called fiducial vectors. This is called ACSQ. There also holds square integrability on $\mathbb{R}_+$ equipped with the measure $dx/x$. Affine coherent states can then be constructed as $\ket{P,R}=U(P,R)\ket{\phi}$, where $\ket{\phi}$ is  fixed but arbitrary. Conditions implemented on $\ket{\phi}$ should be such that numerical values arising in quantum theory are finite. Choosing the fiducial vectors can be seen as a quantization ambiguity in ACSQ, roughly corresponding to the factor ordering problem in Dirac canonical quantization. In practice one usually picks a family of fiducial vectors to see how different choices affect the quantum theory. Following the prescription in this section and modifying \eqref{ciq 1} we get
\begin{equation}
\label{acsq}
    f\rightarrow \Hat{f}=\frac{1}{2\pi \hbar\, c_{-1}^{\phi}}\int_0^{\infty} dR\int_{-\infty}^{\infty} dP   \, f(P,R)\ket{P,R}\bra{P,R},
    \end{equation}
with 
\begin{align}
\label{fiducial1}
     c_{\alpha}^{\phi}&=\int_0^{\infty} \frac{dx}{x^{2+\alpha}} \, \phi^2(x)\\
     I&=\frac{1}{2\pi \hbar\, c_{-1}^{\phi}}\int_0^{\infty} dR\int_{-\infty}^{\infty} dP   \ket{P,R}\bra{P,R}.
\end{align}
An affine coherent state can then be defined as
\begin{equation}
    \ket{P,R}=U(P,R)\ket{\phi}.
\end{equation}
The operator constructed with \eqref{acsq} is symmetric, and if $f$ is at least semibounded it has a self-adjoint extension \cite{Gazeau}. Restricting the lower symbol \eqref{ciq 2} to rank one densities or projection operators, we find the ACSQ version of the lower symbols,
\begin{align}
\label{lower symbol 1}
\Check{f}&=\expval{\Hat{f}}{P,R},\\
\Check{f}&=\frac{1}{2\pi \hbar\, c_{-1}^{\phi}}\int_0^{\infty}dR\int_{-\infty}^{\infty} dP \, f(P',R')\abs{\braket{P',R'}{P,R}}^2.
\label{lower symbol}
\end{align}
In other words, the lower symbol is the expectation value of this operator associated with this function with respect to an Affine Coherent State (ACS). The lower symbol of the Hamiltonian is an important quantity, because it can be used to generate the quantum corrected time evolution in the classical phase space \cite{Bergeron:2013ika}. The viewpoint most commonly adopted seems to be that this quantum corrected sector is not to be treated within a full quantum theory, but rather considered as a theory of its own; for more details, see \cite{tim}, \cite{klaudera}. 
%%%%%%%%%%%%%%%%%%%%%%%%%%%%
\section{Quantum theory for the comoving observer }
\label{comoving obsever qm theory}
 In this section we will apply the ACSQ method to the comoving Hamiltonian found in section \ref{ch:chapter_2},
\begin{equation*}
    H=-\frac{p^2}{6V_sa}-\frac{3V_s}{2}ka.
\end{equation*}
Using the rescaled variables $R=ar_s$, $\frac{p}{r_s}=P$, we arrive at 
\begin{equation}
    H=-\frac{r_s^3}{6V_s}\frac{P_R^2}{R}-\frac{3V_s}{2r_s}kR.
    \label{comoving hamiltonian}
\end{equation}
The Hamiltonian operator in the corresponding quantum theory after ACSQ assumes the form 
\begin{equation}
\begin{aligned}
    \Hat{H}\psi=-\frac{r_s^3\hbar^2}{6V_s c_{-1}^{\phi}}&\left(\frac{1-c_{-4}^{\phi}}{x^2}\psi(x)-\frac{1}{x^2}\psi'(x)+\frac{1}{x}\psi''(x)\right)\\&-\frac{3V_sk c_0^{\phi}}{2r_s c_{-1}^{\phi}} x\psi(x),
    \end{aligned}
    \label{acsq comoving os curved}
\end{equation}
where $r_s$ is the coordinate radius of the OS-model, and  $V_s=\int_0^{r_s}dr\frac{r^2}{\sqrt{1-kr^2}}$. Technical details for the ACSQ quantization can be found in appendix \ref{ACSQ calculation for comoving OS observer}. Using equations \eqref{acsq os finale} and \eqref{R Finale}, we find \eqref{acsq comoving os curved}. Now that we have $\Hat{H}$ let us calculate the lower symbol. As discussed in section \ref{acsq section}, the lower symbol function is central in the ACSQ method, since it provides us with a way to perform a semiclassical analysis. Using equations \eqref{lower symbol comoving os finale app} and \eqref{lower symbol r^beta finale}, we find for the lower symbol: 
\begin{equation}
\begin{aligned}
    \Check{H}=&-\left(\frac{r_s^3}{6V_s}\right)\left(\frac{P^2}{R}+\frac{ c_{-1}^{\phi'}+c_1^{\phi}{\left(c_{-4}^{\phi'}-1\right)}}{c_{-1}^{\phi}}\frac{\hbar^2}{R^3}\right)\\&-\frac{3V_sk}{2r_s} \frac{c_{-3}^{\phi}c_0^{\phi}}{c_{-1}^{\phi}}R.
    \end{aligned}
\end{equation}
Identifying 
\begin{align*}
    \delta_0=&\frac{c_{-1}^{\phi'}+c_{1}^{\phi}\left(c_{-4}^{\phi'}-1\right)}{c_{-1}^{\phi}},\\ \delta_1=&\frac{c_{-3}^{\phi}c_0^{\phi}}{c_{-1}^{\phi}},
\end{align*}
we get
\begin{equation}
\begin{aligned}
    \Check{H}=&-\left(\frac{r_s^3}{6V_s}\right)\left(\frac{P^2}{R}+\frac{\hbar^2 \delta_0}{R^3}\right)-\frac{3V_s\,k}{2r_s}\delta_1 R\\=&-\frac{3V_s}{r_s^3}\Bar{M}
    \label{comoving manipulation 0}.
    \end{aligned}
\end{equation}

This Hamiltonian matches the one for the Friedmann models found in \cite{klaudera} using Klauder's enhanced quantization. Let us recall that we identified our Hamiltonian for the comoving observer as $H=-\frac{3V_s}{r_s^3}M$, which insures that energy is non-negative. Using Hamilton equations we have, on the one hand, 
\begin{align}
    \pdv{H}{P}=&\Dot{R},\\
    -\frac{r_s^3}{3V_s}\frac{P}{R}=&\Dot{R}.
    \label{comoving manipulations 1}
    \end{align}
On the other hand, rearranging \eqref{comoving manipulation 0} we find 
\begin{equation}
    \frac{r_s^6}{18V_s}\frac{P^2}{R}={M}-\frac{r_s^6}{18V_s}\frac{\hbar^2 \delta_0}{R^3}-\frac{3kr_s^2}{2}\delta_1R.
    \label{comoving manipulation 2}
\end{equation}
Dividing \eqref{comoving manipulation 2} by $R$ we arrive at
\begin{equation}
    \frac{r_s^6}{18V_s}\frac{P^2}{R^2}=\frac{M}{R}-\frac{r_s^6}{18V_s}\frac{\hbar^2 \delta_0}{R^4}-\frac{3kr_s^2}{2}\delta_1.
    \label{comoving manipulation 3}
\end{equation}
In the next step we take the square of \eqref{comoving manipulations 1} and compare it with \eqref{comoving manipulation 3} to find
\begin{equation}
    \Dot{R}^2=\frac{2M}{R}-\frac{r_s^6}{9V_s}\frac{\hbar^2\delta_0}{R^4}-3kr_s^2 \delta_1,
    \label{graph1}
\end{equation}
where the dot denotes the time derivative with respect to proper time. We define
\begin{align*}
    \alpha=&\frac{r_s^6\hbar^2}{9V_s},\\
    \beta=&\frac{3kr_s^2}{2}
\end{align*}
and write
\begin{equation}
   \pdv{R}{\tau}=\frac{\sqrt{2MR^3-\alpha\delta_0-\beta R^4 \delta_1}}{R^2}.
\end{equation}
From the last equation we then find
\begin{equation}
    \int d\tau=\int_0^{\infty}\frac{dRR^2}{\sqrt{MR^3-\alpha \delta_0-\beta \delta_1 R^4}}.
    \label{comoving os manipulation 4}
\end{equation}
Unfortunately, this integral does not converge. But if we look at the flat case, which was investigated in \cite{tim}, we can set $\beta=0$ and find the following solution for $R(\tau)$:
\begin{equation}
    R(\tau)=\left(\frac{\hbar^2 \delta_0}{M}+\frac{9M}{2}\left(\tau-\tau_0\right)^2\right)^{\frac{1}{2}}.
\end{equation}
This is an important result, since it exactly matches the quantum corrected dust trajectories found in \cite{LTB}. It confirms that the ACSQ method leads to reliable results. In this case it exactly reproduces the result found in \cite{LTB} within the Dirac quantization scheme. For large $R$ the trajectory reproduces the classical expanding and collapsing trajectories, which are then connected by a bounce that replaces the collapse to a singularity. The lifetime can be  estimated by 
\begin{equation}
    \Delta \tau=\tau_+-\tau_-,
\end{equation}
where $R(\pm\tau)=2M$. In order to calculate this lifetime we consider a comoving observer on the Oppenheimer-Snyder radius  who observes the matter that collapses and bounces back. One finds the following expression for the lifetime: 
\begin{equation}
    \Delta\tau=\frac{8M}{3}\sqrt{1-\left(\frac{R_0}{2M}\right)^3}.
\end{equation}
From the viewpoint of the comoving observer,
the lifetime is thus proportional to the mass of the dust cloud. Moreover, when $R_0\ll2M$ one recovers the result derived for an comoving observer in \cite{LTB}
within the Dirac quantization scheme.

\begin{figure}[h!]
    \centering
    \includegraphics[width=55mm]{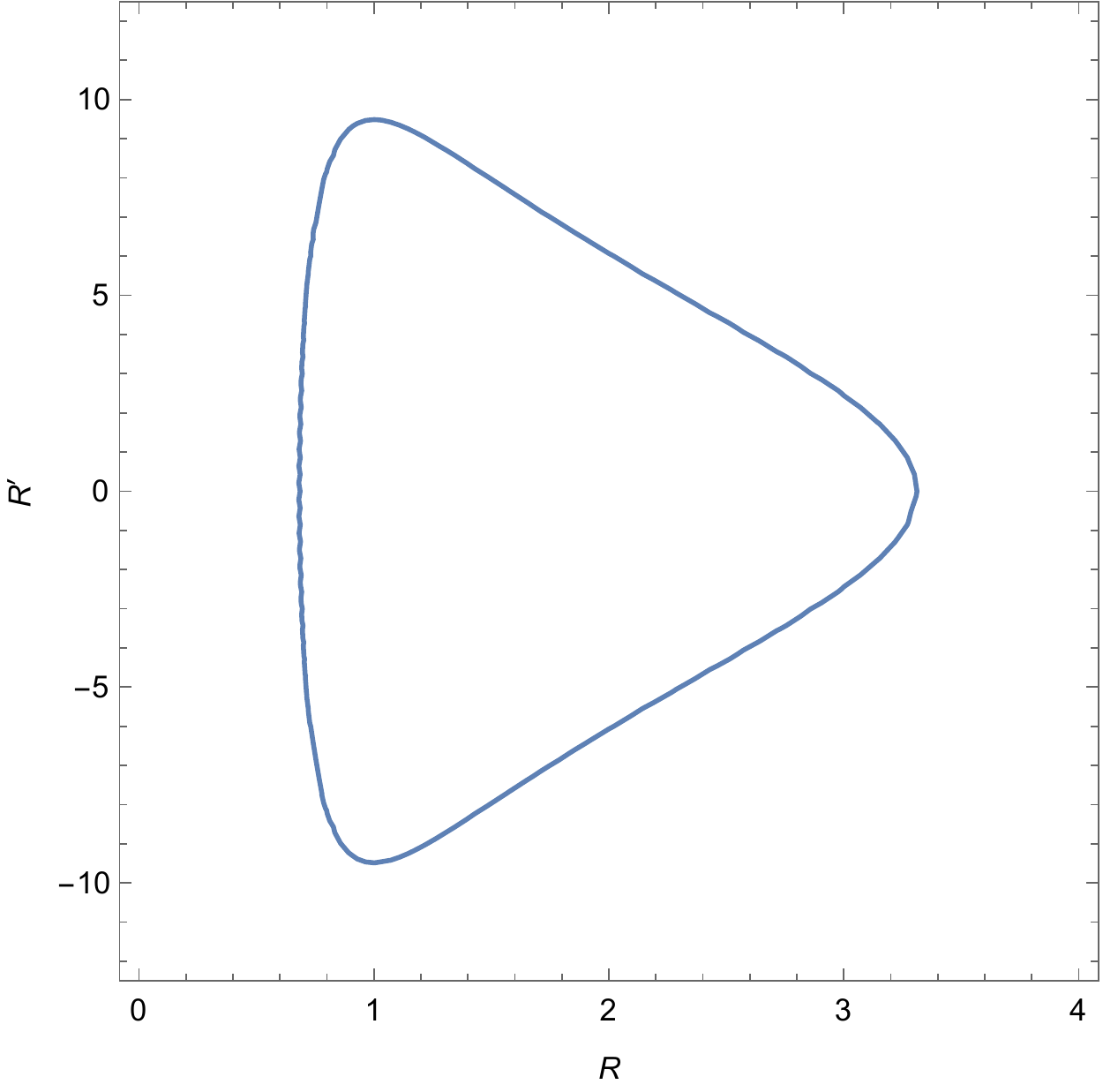}
    \caption{Quantum corrected phase space portrait for the comoving observer and closed space-time in the $\Dot{R}-R$ plane, with numerical values $k=1,M=100,b=50,a=60$. Here, and in the figures below, $M$ is written in units of the Planck mass.}
    \label{fig:comoving closed}
\end{figure}

We note that after implementing ACSQ we simply get the classical Hamiltonian plus a potential term. This potential can be shown in the flat case to be always repulsive. Therefore we have a simple bounce that replaces the singularity. But the situation is not so simple in the closed case, since it has another potential term proportional to $R$. Unfortunately, because \eqref{comoving os manipulation 4} does not posses an analytical solution, we cannot perform the same analysis discussed here for the closed case. Therefore in order to perform a semiclassical analysis we resort to Mathematica and plot the phase space by using the lower symbols in ACSQ. Then we try to analyze the plot to learn more about the closed case. Plotting \eqref{graph1} by identifying 
\begin{align*}
    a=&3kr_s^2 \delta_1\\
    b=&\frac{r_s^6 \hbar^2 \delta_0}{9V_s },
\end{align*}
we find figure \ref{fig:comoving closed}.
Now that we have the quantum corrected phase space for the closed case, we can plot the quantum corrected phase space for the flat case and compare the two cases.
The flat case implies $a=0$.
\begin{figure}[h!]
    \centering
    \includegraphics[width=55mm]{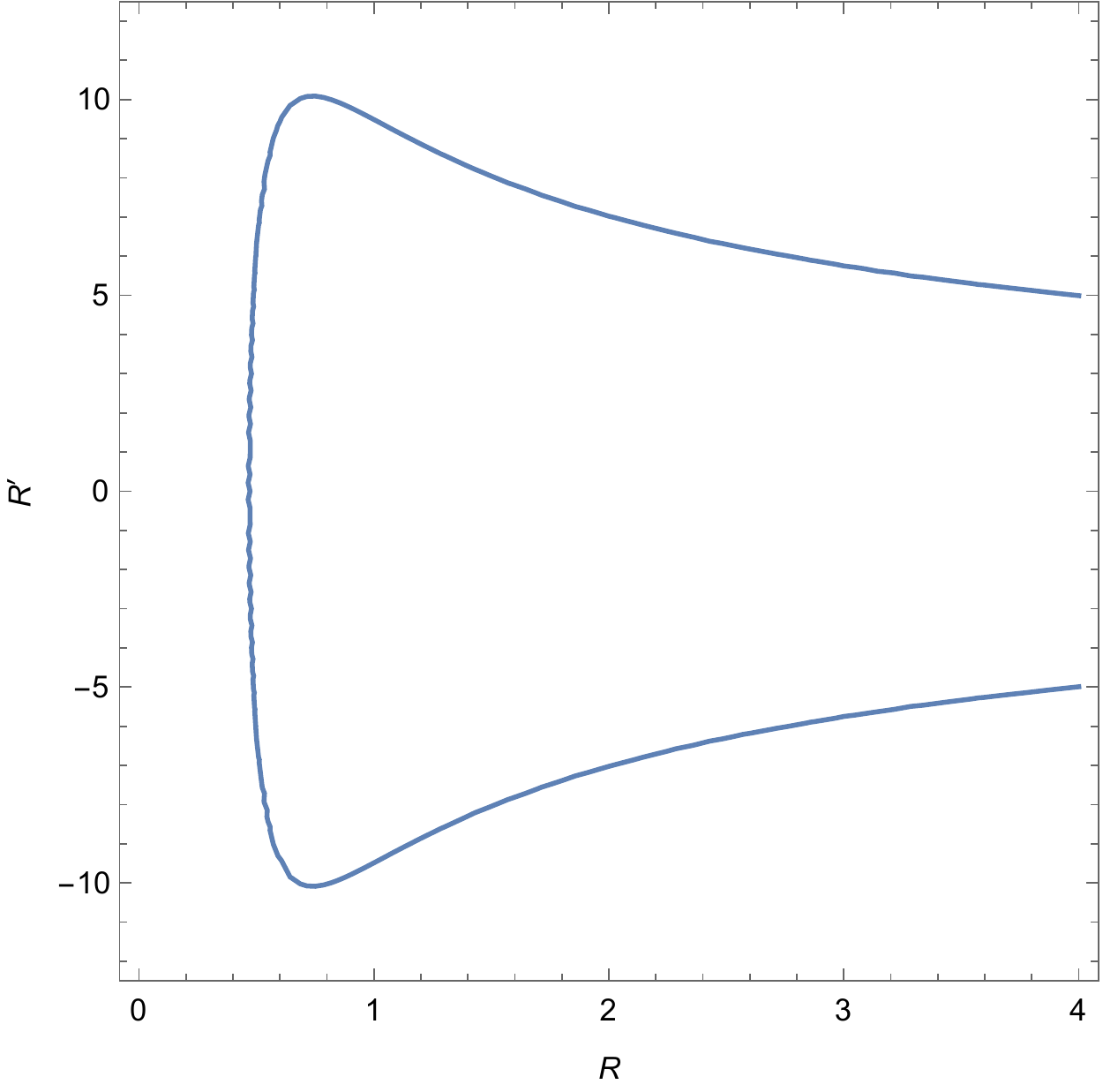}
    \caption{Comoving quantum corrected phase space portrait for the flat case in $\Dot{R}-R$ plane, with numerical values $a=0,b=10,m=50$. }
    \label{fig:comoving flat}
\end{figure}
Comparing figures \ref{fig:comoving closed} and \ref{fig:comoving flat} exhibits that in the closed case we have an oscillatory behavior. The matter falls towards the singularity and bounces; after the bounce it recollapses, different from the flat case which just has one bounce. We have to emphasize that a bounce is not a generic feature of these models. It depends on how we choose our $c_{\alpha}^{\phi}, \delta_0, \delta_1,M$; for example, by choosing $a<b$, bounce and collapse will not happen and we get back the figure for the flat case. Furthermore, $M$ has to be significantly greater than $a$; otherwise we will not get a bounce and recollapse behavior. Recall that the horizon is at $R=2M$, so the discussion in figure \ref{fig:comoving closed} and \ref{fig:comoving flat} is located well within the horizon. This is one of the advantages of investigating a comoving observer, since it gives us the ability to look past through the horizon at least in theory and provides us with a better understanding of the bigger picture. We have established so far that there is an oscillatory behavior between bounce and recollapse; 
we claim that all these oscillations are well within the horizon. 

\begin{figure}[h!]
    \centering
    \includegraphics [width=60mm]{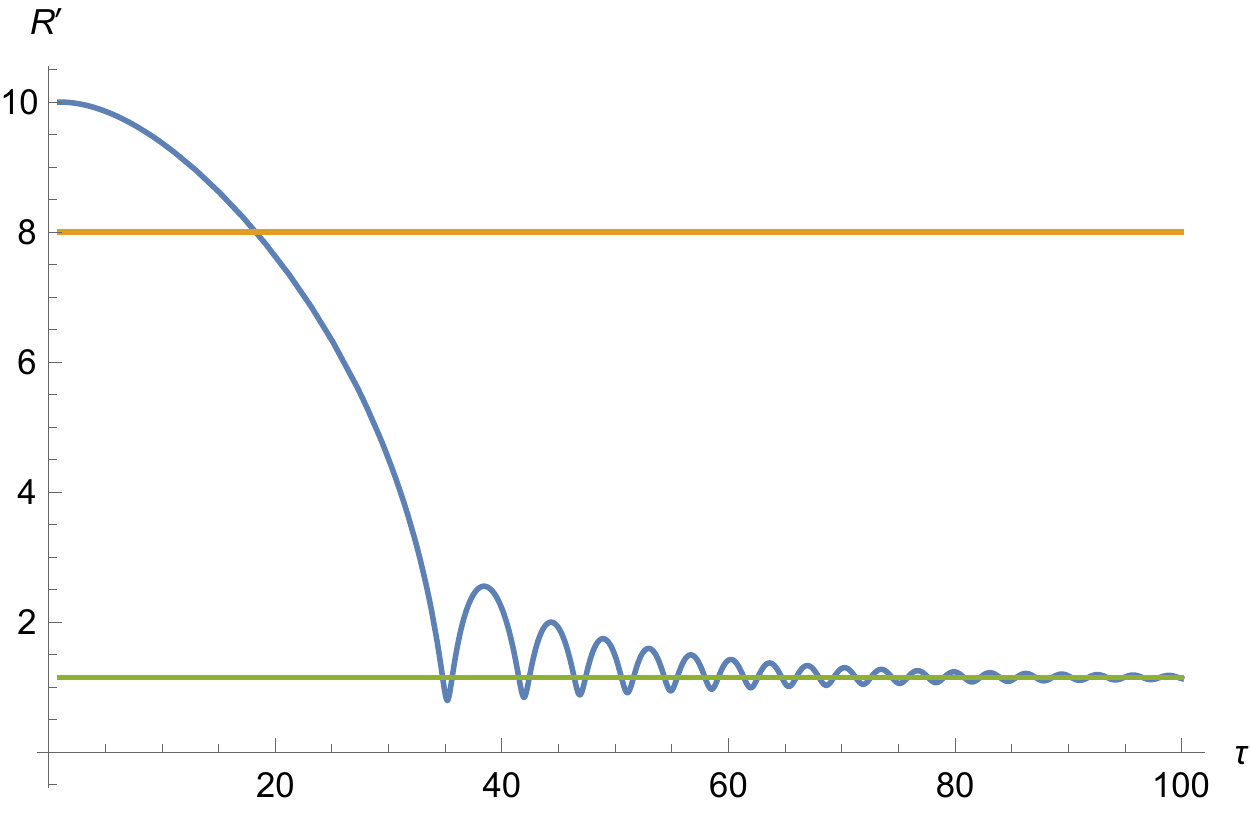}
    \caption{Graph in $R-\tau$ space with numerical values $M=4, b=6, c=0.1$. The orange line represents the horizon $R=2M$, the green line represents the location where this model reaches equilibrium after oscillations, where $\Dot{R}=\Ddot{R}=0$. }
    \label{fig:damped blackhole}
\end{figure}

Naturally the question arises how robust this claim is and how heavily it depend on the parameters. In order to investigate this we differentiate \eqref{graph1} with respect to $\tau$ (dust proper time) and add a small perturbation to the differential equation in the form of $-c\Dot{R}$, where $0<c\ll1$ is a numerical constant, to get
\begin{align}
    \Dot{R}\Ddot{R}&=-\frac{M}{R^2}\Dot{R}+\frac{2r_s^6\hbar^2\delta_0}{9V_sR^5}\Dot{R},\\
    \Ddot{R}&=-\frac{M}{R^2}+\frac{2r_s^6\hbar^2\delta_0}{9V_sR^5}-c\Dot{R}.
    \label{damped blackhole}
\end{align}
Unfortunately, the differential equation \eqref{damped blackhole} does not have an analytical solution, which is why we present here a numerical solution using Python and produce a plot of $R$ with respect to dust proper time $\tau$ in figure \ref{fig:damped blackhole}. The wiggly line can be interpreted as oscillations representing quantum fluctuations. Our analysis interestingly shows that these oscillations never spill outside of the horizon for different numerical values in the phase space. This can be interpreted as if looking from the outside, this object is behaving like a classical Schwarzschild black hole since it mimicks an eternal lifetime. There is thus a consistency between the finite lifetime and the apparent eternal lifetime of the classical situation. We also note that there are some fast oscillations in figure \ref{fig:damped blackhole} and that the system reaches equilibrium (green line) quite rapidly.

%%%%%%%%%%%%%%%%%%%%%%%%%%%%%%%%%%%%%%%
\section{Quantum theory for the stationary observer}
\label{quantum theory for the stationary observer}
After having discussed in the last section the case of a {\em comoving} observer, we now turn to the case of a {\em stationary} observer sitting at rest far away from the dust cloud. It is this case for which the advantages of the ACSQ method are most clearly seen.
 
From the point of view of such an observer, we have found 
\eqref{stationary hamiltonian finale 4}.
%\begin{align*}
%    {H}=\frac{3\Bar{R}\,V_s}{2\sqrt{p_k}r_s^3 p_k} \, \begin{cases}
 %   1-p_k-\tanh^2{\left(\frac{r_s^3\Bar{P}_{\Bar{R}}}{3V_s\Bar{R} }\sqrt{p_k}\right)}\\
%     1-p_k-\coth^2{\left(\frac{r_s^3\Bar{P}_{\Bar{R}}}{3V_s\Bar{R} }\sqrt{p_k}\right)}
%    \end{cases}.
 %   \end{align*}
For quantization, it is convenient to perform a canonical transformation and bring this Hamiltonian into a slightly different form. The transformation reads 
\begin{align}
\label{statinery canonical 1}
    P_A=&\frac{P_R}{R}\frac{r_s^3 \sqrt{p_k}}{3V_s},\\
    A=&\frac{3R^2V_s}{2r_s^3\sqrt{p_k}}.
    \label{stationery canonical 2}
\end{align}
To see explicitly that it is a canonical transformation, we check that the Poisson bracket remains unchanged,
\begin{align}
    \qty\bigg{P_A,A}=\pdv{P_A}{P_R}\pdv{A}{R}-\pdv{A}{P_R}\pdv{P_A}{R}=1.
\end{align}
Rewriting the Hamiltonian \eqref{stationary hamiltonian finale 4} by using \eqref{statinery canonical 1} and \eqref{stationery canonical 2}, we get 
\begin{align}
    {H}=\frac{\sqrt{3V_s}\sqrt{A}}{r_s^{\frac{3}{2}}p_k^{\frac{5}{4}}\sqrt{2}} \, \begin{cases}
    1-p_k-\tanh^2{P_A}\\
     1-p_k-\coth^2{P_A}
    \end{cases}.
    \label{stationary hamiltonian chapter 6}
\end{align}
 If we want to directly quantize \eqref{stationary hamiltonian chapter 6}, we have to allow multivalued quantum states, with both branches evolving with respect to a branch of the Hamilton operator. In \cite{saphere} it is shown that the Hamiltonian as a function of momentum is a branched function with cusps; appropriate boundary conditions are there identified to insure unitary evolution. Strictly speaking, in such a multivalued quantum theory, unitary evolution can still be implemented with boundary conditions on the wave function at Hamiltonian branching points. Because we are mainly interested in the quantum corrected dynamics of the system, not the wave function directly, we will not exactly implement the construction from \cite{saphere}. The branching points for our Hamiltonian are at $P_A\rightarrow \infty$; one can interpret this as quantization of two different cases, as discussed for the flat case in \cite{tim}. Also our analysis of quantum corrected trajectories demonstrates that we do not need to explicitly construct multivalued quantum states at this level. It appears that the branching point of the Hamiltonian $P_A\rightarrow \pm \infty$ can never be reached in finite time. 
 
 Nevertheless,
there are different approaches in quantizing multivalued systems; in \cite{PhysRevA.36.4417}, for example, an effective Hamiltonian was found as a sum of the different branches of the original multivalued one by considering the path integral. This method relies on a multivalued Hamiltonian coming from a  single-valued Lagrangian where velocities usually occur with power higher than two. This is not useful for our Hamiltonian, and therefore we stay close to the approach developed in \cite{saphere}. In section \ref{acsq section} we have discussed that every complicated phase space function can be associated with an operator; if the function is at least semibounded, it will have a self-adjoint extension. Therefore the fact that hyperbolic tangent and cotangent do not have Taylor expansions everywhere is not a problem here. Note that in our Hamiltonian both hyperbolic tangent and cotangent are semibounded, so the resulting operator is self-adjoint and the complicated dependence on the momentum is not a problem as well. But the Hamilton operators acting in position space are the sum of a multiplicative and unbounded integral operator. Therefore it is very difficult to find eigenfunctions and eigenvalues, if not impossible, which is why we shall concentrate on quantum corrections using lower symbols here. 

Let us therefore
calculate the lower symbol of \eqref{stationary hamiltonian chapter 6}. We start by using equation \eqref{lower symbol}; we shall denote the outside branch by $H_+(P_A,A)$ and the inside branch by$H_-(P_A,A)$. Their respective lower symbols read 
\begin{equation}
    \begin{aligned}
        \check{H}_+=&-\frac{1}{2\pi\hbar c_{-1}^{\phi}}\int_0^{\infty}d\Bar{A}\int_{-\infty}^{\infty} d\Bar{P}_A\abs{\bra{\Bar{P}_A,\Bar{A}}\ket{{P}_A,{A}}}^2\\& \left[-\alpha(1-p_k)\sqrt{\frac{A}{2}}-\alpha\sqrt{\frac{A}{2}}\left(\tanh^2{\bar{P}_A}-1\right)-\alpha\sqrt{\frac{A}{2}}\right],
        \label{outside branch}
    \end{aligned}
\end{equation}
and 
\begin{equation}
    \begin{aligned}
        \check{H}_-=&-\frac{1}{2\pi\hbar c_{-1}^{\phi}}\int_0^{\infty}d\Bar{A}\int_{-\infty}^{\infty} d\Bar{P}_A\abs{\bra{\Bar{P}_A,\Bar{A}}\ket{\Bar{P}_A,\Bar{A}}}^2\\& \left[-\alpha(1-p_k)\sqrt{\frac{A}{2}}-\alpha\sqrt{\frac{A}{2}}\left(\coth^2{\bar{P}_A}-1\right)-\alpha\sqrt{\frac{A}{2}}\right],
        \label{inside branch}
    \end{aligned}
\end{equation}
with $\alpha_1=\sqrt{3V_s}/r_s^{\frac{3}{2}}p_k^{\frac{5}{4}}$. We have slightly rewritten the Hamiltonian in order to simplify our calculations by using the identities
\begin{equation}
\begin{aligned}
     &\int_{-\infty}^{\infty} d\Bar{P}_A\left(\tanh^2{\Bar{P}_A}-1\right)e^{-\frac{i}{\hbar}\Bar{P}_A(x-y)}\\&=-\frac{\frac{\pi}{\hbar}(x-y)}{\sinh{\frac{\pi}{2\hbar}(x-y)}},
     \label{identity1}
     \end{aligned}
\end{equation}
\begin{equation}
\begin{aligned}
    & \int_{-\infty}^{\infty} d\Bar{P}_A\left(\coth^2{\Bar{P}_A}-1\right)e^{-\frac{i}{\hbar}\Bar{P}_A(x-y)}\\&=-\frac{\frac{\pi}{\hbar}(x-y)}{\tanh{\frac{\pi}{2\hbar}(x-y)}},
     \label{identity 2}
     \end{aligned}
\end{equation}
where
\begin{equation}
\begin{aligned}
\abs{\bra{\Bar{P}_A,\Bar{A}}\ket{P_A,A}}^2=&\int_0^{\infty}dx\int_0^{\infty}dy \frac{e^{\frac{i}{\hbar}P_A-\Bar{P}_A}(x-y)}{A\Bar{A}}\\ &\phi(\frac{x}{A}) \phi^*(\frac{x}{\Bar{A}}) \phi^*(\frac{y}{A})\phi(\frac{y}{\Bar{A}}).
\end{aligned}
\end{equation}
The first Fourier transform \eqref{identity1} can be evaluated by contour integration. For computing the second Fourier transform \eqref{identity 2} one has to be more careful because the integrand diverges as $\Bar{P}_A\rightarrow 0$.
Nevertheless, one can regularize this by performing two contour integrations, one over the contour including the divergence and one excluding it, and averaging over the results. After straightforward but tedious calculations the lower symbols take the form  
\begin{equation}
    \begin{aligned}
        H_{\pm}=&-\alpha_1\left(p_k-2\right)\frac{c_{-\frac{5}{2}}^{\phi}c_{-\frac{1}{2}}^{\phi}}{c_{-1}^{\phi}}\sqrt{\frac{A}{2}}\\&+\frac{\alpha_1}{2\sqrt{2}\hbar^2 c_{-1}^{\phi}A} \int_0^{\infty}\frac{d\Bar{A}}{\sqrt{\Bar{A}}}\int_0^{\infty}dx\\ &\int_0^{\infty}dy\,\frac{x-y}{F_{\pm}(\frac{\pi}{2\hbar}(x-y))}e^{\frac{i}{\hbar}P_A(x-y)}\\ &\phi(\frac{x}{A})\phi^*(\frac{y}{A})\phi^*(\frac{x}{\Bar{A}})\phi(\frac{y}{\Bar{A}})
   , \end{aligned}
   \label{primary lower symbol}
\end{equation}
where 
\begin{align*}
    F_+=&\sinh(x)\\
    F_-=&\tanh(x).
\end{align*}
For details regarding contour integration for \eqref{identity1} and \eqref{identity 2}, we refer to \cite{tim}. 
To continue we need to specify a fiducial vector. A convenient choice is made in \cite{klaudera},
\begin{equation}
    \phi(x)=\frac{{(2\beta)^{\beta}}}{\sqrt{\Gamma(2\beta)}}x^{\beta- \frac{1}{2}}
    \,e^{-\beta x},
\end{equation}
\begin{align}
    c_{-1}^{\beta}=&\frac{2\beta}{2\beta-1}\\
    c_{-\frac{1}{2}}^{\phi}=&2\sqrt{2}\,\beta^{\frac{3}{2}}\frac{\Gamma(2\beta-\frac{3}{2})}{\Gamma(2\beta)}\\c_{-\frac{5}{2}}^{\phi}&=\frac{\Gamma(2\beta+1)}{\sqrt{2\beta}\Gamma(2\beta)}\\c_{-3}^{\phi}&=1.
\end{align}
In order to insure that these constants are finite, we impose $\beta>\frac{3}{4}$. We then perform the$\Bar{A}$ integration using the identity 
\begin{equation*}
    \int_0^{\infty} x^{t-1}e^{-x}dx=\Gamma(t),
\end{equation*}
and find 
\begin{equation}
    \begin{aligned}
        \int_0^{\infty} \frac{d\Bar{A}}{\sqrt{\Bar{A}}}&\phi^*(\frac{x}{A})\phi(\frac{y}{A})\\&=\frac{(2\beta)^{\beta}}{\Gamma(2\beta)}(xy)^{\beta-\frac{1}{2}}\int_0^{\infty}\frac{d\Bar{A}}{A^{-2\beta-\frac{1}{2}}} e^{-\frac{\beta}{A}(x+y)}\\&=\frac{2^{2\beta}\beta^{\frac{3}{2}}}{\Gamma(2\beta)}\frac{(xy)^{\beta-\frac{1}{2}}}{(x+y)^{2\beta-\frac{3}{2}}}\Gamma(2\beta-\frac{3}{2}),
    \end{aligned}
\end{equation}
so
\begin{equation}
    \begin{aligned}
        &\frac{\Gamma(2\beta-1)\Gamma(2\beta)}{\Gamma(2\beta-\frac{3}{2})\Gamma(2\beta+\frac{1}{2})}\check{H}_{\pm}=\\& \alpha_1(p_k-2)\sqrt{\frac{A}{2}}\\&+\frac{2^{4\beta-\frac{5}{2}}\beta^{2\beta+\frac{1}{2}}\alpha_1}{\hbar^2\Gamma(2\beta+\frac{1}{2})A^{2\beta}}\int_0^{\infty}dx\\ &\int_0^{\infty}dy\frac{x-y\,(xy)^{2\beta-1}}{F_{\pm}(\frac{\pi}{2\hbar}(x-y))(x+y)^{2\beta-\frac{3}{2}}}e^{\frac{i}{\hbar}P_A(x-y)-\frac{\beta}{A}(x+y)}.
    \end{aligned}
    \label{stationary intermediate}
\end{equation}
In addition, we know from \eqref{dust cloud mass 4.} \eqref{76} that $H_{\pm}=3V_s M/r_s^3 \sqrt{p_k}$. 
This yields the expression \eqref{stationery finale 6} for the mass of the dust cloud : 
\begin{equation}
    M=- \frac{\Gamma(2\beta-1)\Gamma(2\beta)}{\Gamma(2\beta-\frac{3}{2})\Gamma(2\beta+\frac{1}{2})} \check{H}_{\pm}(P_A,A)\frac{r_s^3}{3V_s}\sqrt{p_k}\,.
    \label{stationery finale 6}
\end{equation}

\begin{figure}[h!]
    \centering
    \includegraphics[width=60mm]{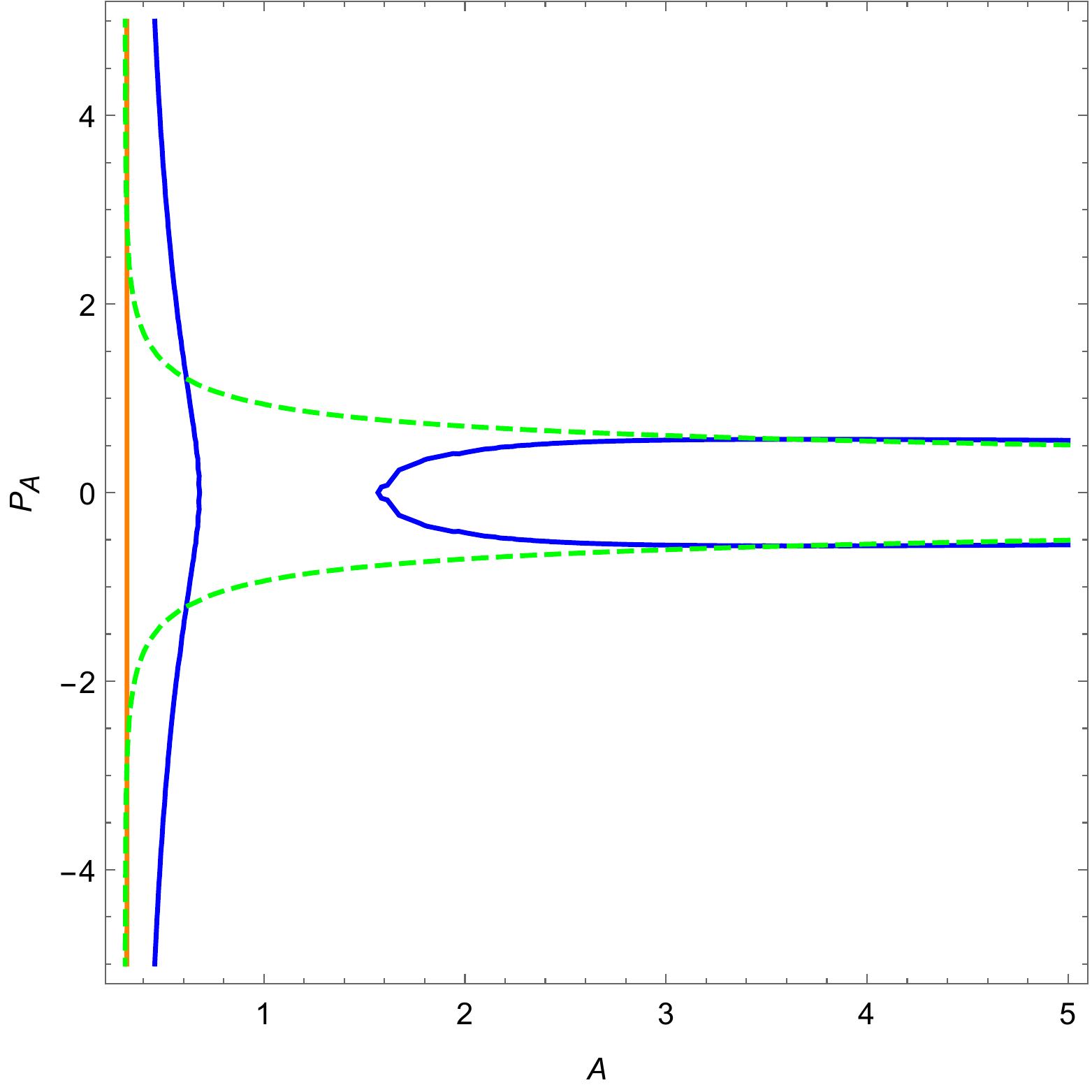}
    \caption{Quantum corrected phase space portraits for the outside branch of the Hamiltonian given by the blue line as compared to the classical counterpart given by the dashed green line. The orange line determines the horizon. The parameter choices are $\beta=5,M=1,r_s=0.2$}
    \label{fig:stationeryoutside branch  bounce}
\end{figure}

\begin{figure}[h!]
    \centering
    \includegraphics[width=60mm]{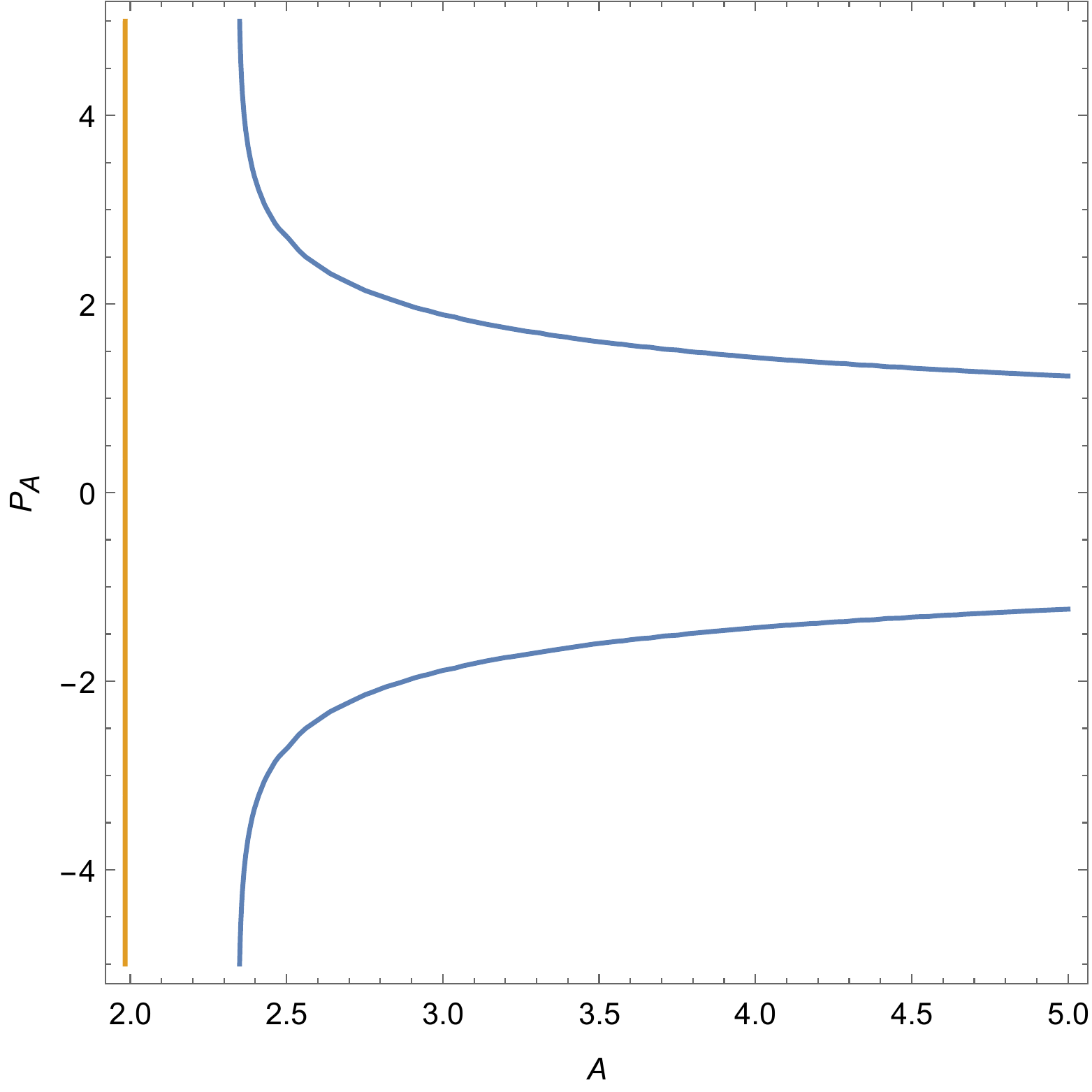}
    \caption{Quantum corrected phase space for the outside branch when $r_s=0.2,  \beta=1, M=1$; the blue line is the quantum corrected trajectory and the orange line is the horizon.}
    \label{fig:outside branch qm classic behavior}
\end{figure}

\begin{figure}[h!]
    \centering
    \includegraphics[width=55mm]{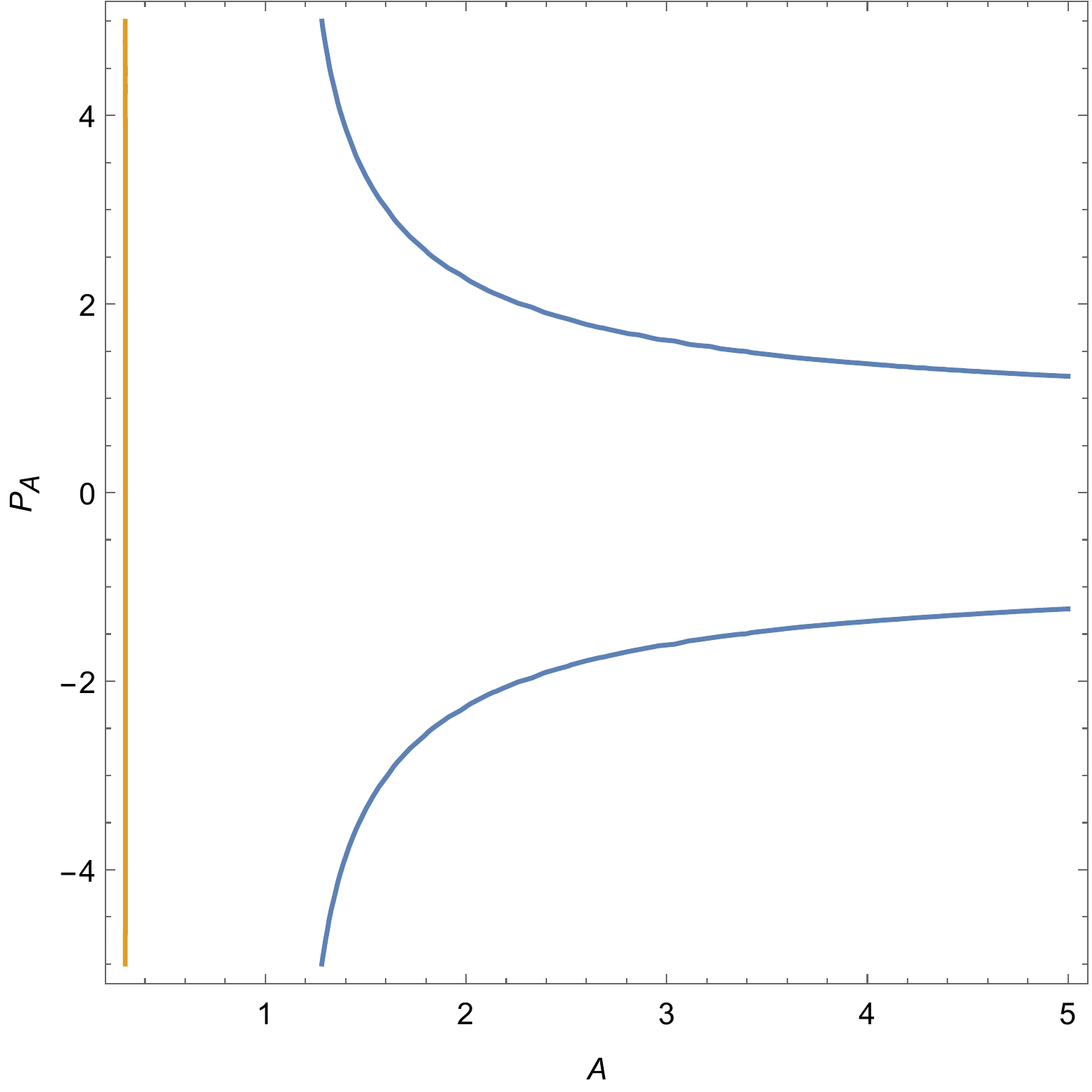}
    \caption{Quantum corrected portrait for the outside branch with $r_s=0.5,\beta=5,M=0.4$ Blue line is quantum corrected trajectory and the orange line is the horizon.}
    \label{fig:classical rs big}
\end{figure}

\begin{figure}[h!]
    \centering
    \includegraphics[width=60mm]{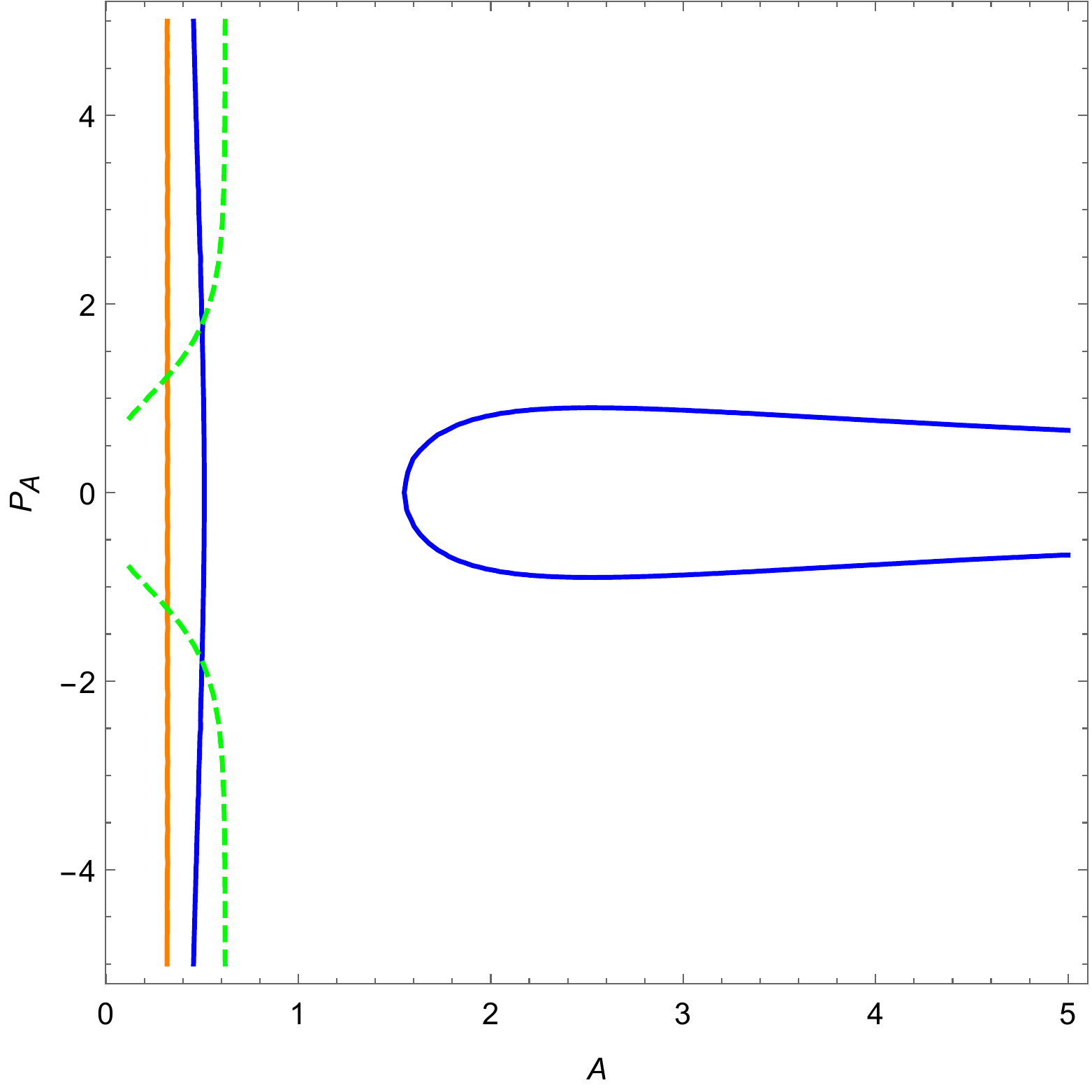}
    \caption{Quantum corrected portrait for the inside branch which resembles a bounce for $\beta=10,M=0.4,r_s=0.2$. The orange line denotes the horizon $R=2M$, the blue line shows the bounce and collapse behavior of the inside branch, and the dashed green line is the classical portrait. }
    \label{fig:inside branch bounce}
\end{figure}

\begin{figure}[h!]
    \centering
    \includegraphics[width=60mm]{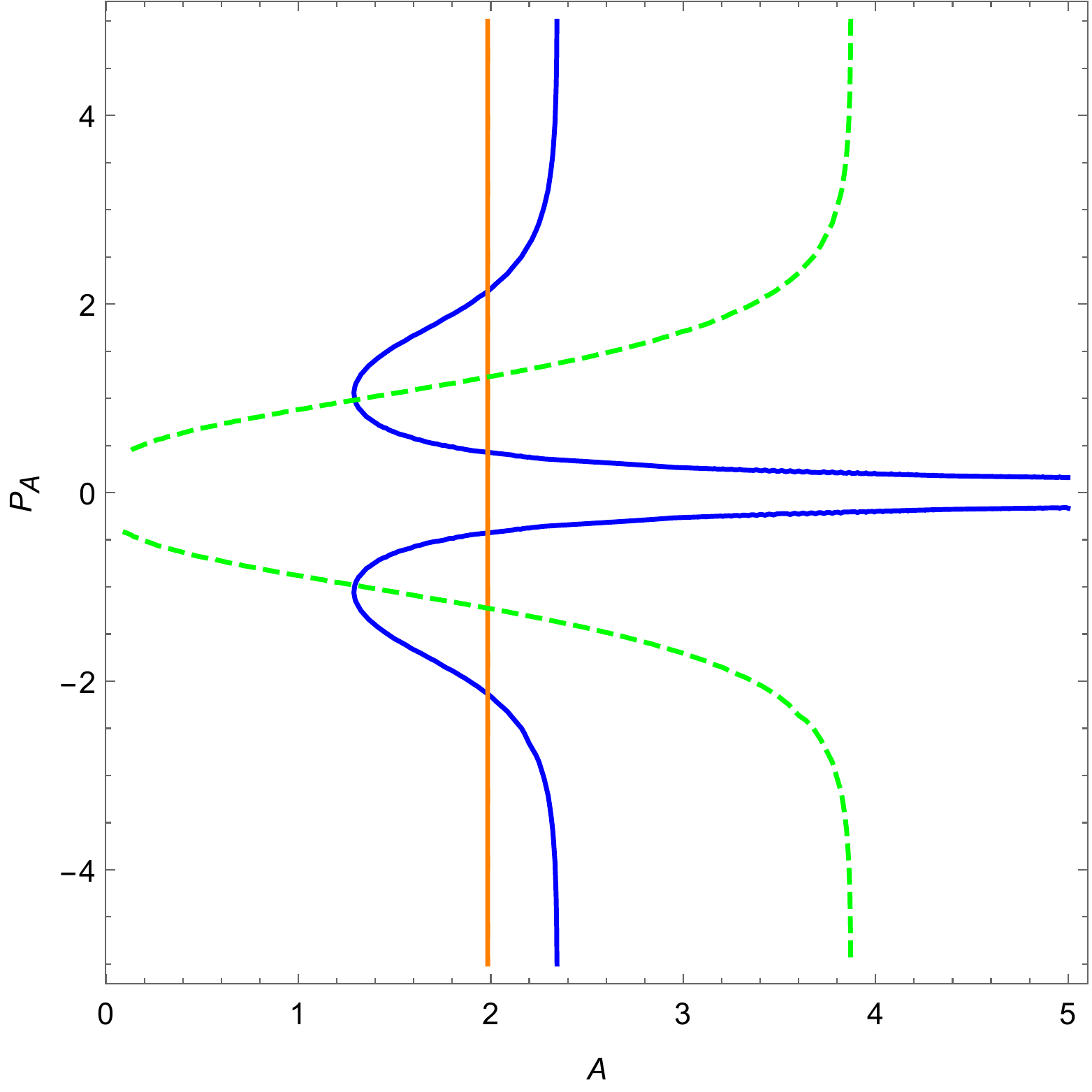}
    \caption{Quantum corrected portrait of the inside branch $\beta=M=1,r_s=0.2$. The blue line shows the quantum corrected inside branch, the orange line the horizon, and the dashed line is the classical trajectory.}
    \label{fig:inside not bounce}
\end{figure}

\begin{figure}[h!]
    \centering
    \includegraphics[width=60mm]{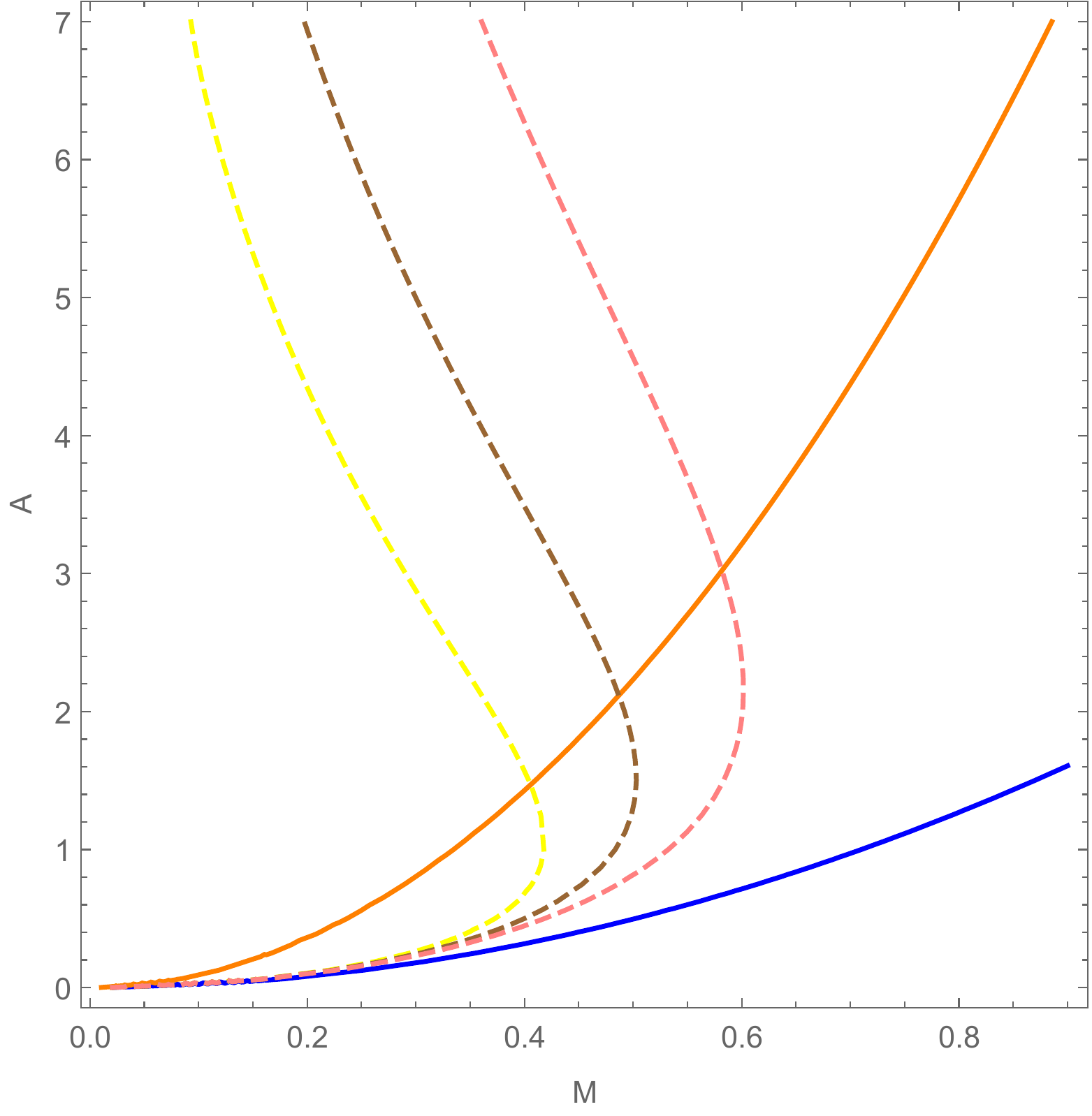}
    \caption{Quantum corrected phase-space portrait at $P_A=0,r_s=0.2$ for the outside branch of the Hamiltonian given by \eqref{stationery finale 6}; $\beta=5,10,20$ are given by dashed lines, the photon sphere is the full orange line $R=3M$, and the horizon the full blue line $R=2M$. }
    \label{fig:approximation rs 0.2}
\end{figure}

\begin{figure}[h!]
    \centering
    \includegraphics[width=40mm]{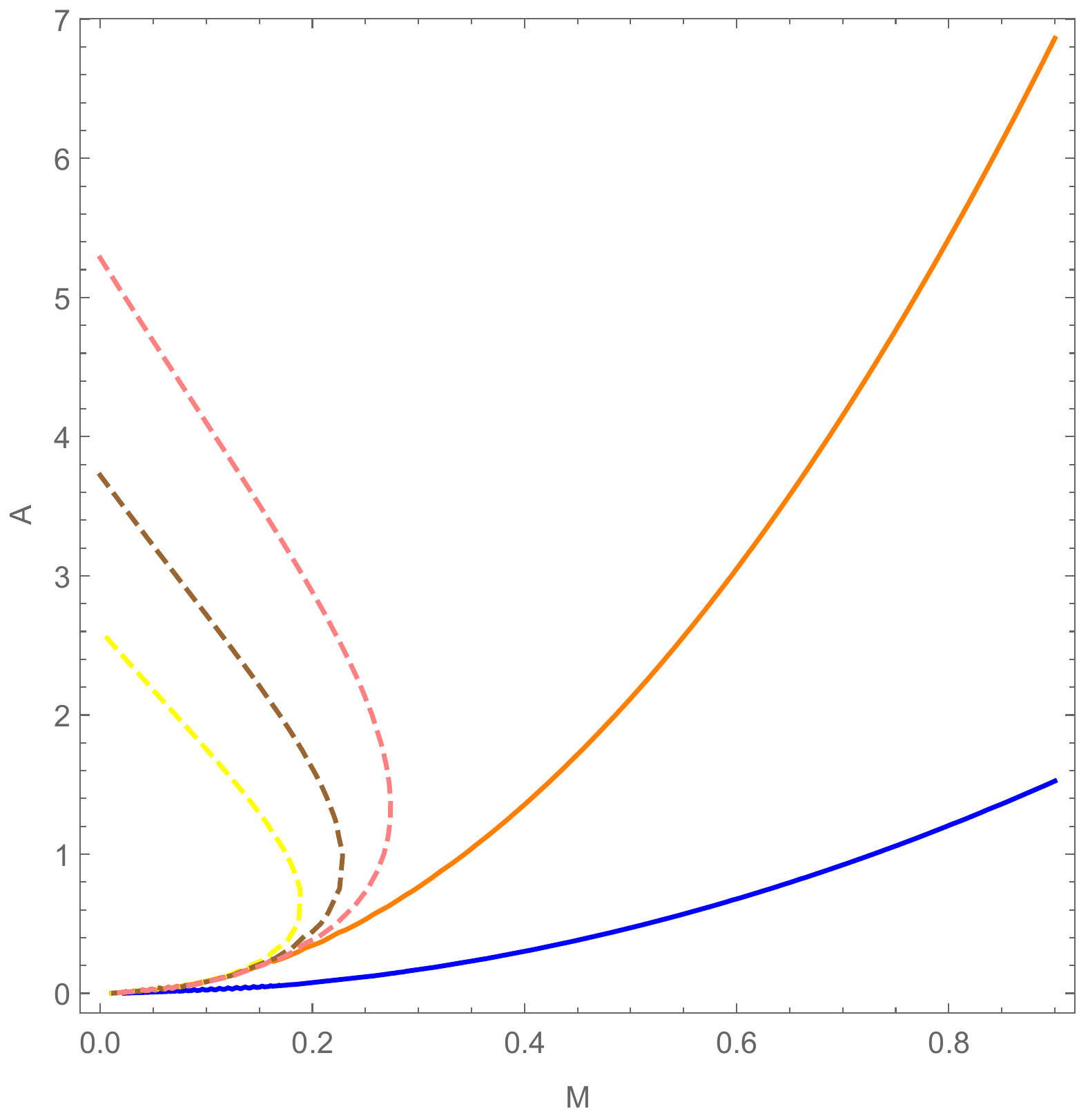}
    \caption{Quantum corrected phase-space portrait at $P_A=0,r_s=0.5$ for the outside branch of the Hamiltonian as given by \eqref{stationery finale 6}. The values $\beta=5,10,20$  are given by dashed lines, the photon sphere by the full orange line $R=3M$, and the horizon by the full blue line $R=2M$.}
    \label{PA 0 rs 0.5}
\end{figure}

The above quantum corrected Hamiltonians can be investigated numerically. For this, we present the phase space portraits and analyze them subsequently. 
We show the results in figures \ref{fig:stationeryoutside branch  bounce} to \ref{PA 0 rs 0.5}.

We note that the graphs shown here are restricted to masses close to the Planck mass. 

For $\beta=M=1$ (in Planck units) and $r_s=0.2$, the outside branch phase space portrait is given by figure \ref{fig:outside branch qm classic behavior}; a similar behavior is found for $\beta<M$. Kinematically, one
can divide the system into two parts, one asymptotically collapsing towards
the horizon and one expanding away from it. When one chooses higher $\beta\gg M$
and $r_s=0.2$, the branches collapsing from and 
expanding to infinity become connected, see figure \ref{fig:stationeryoutside branch  bounce}. This simply resembles itself as the bounce of the dust cloud when collapsing from infinity and recollapse when expanding away from the horizon. Increasing the coordinate radius of the model to $r_s>0.4$ would fully destroy the bouncing collapse picture and would force the system to behave like the classical one, see figures \ref{fig:stationeryoutside branch  bounce} and \ref{fig:classical rs big}. One could possibly compensate for increasing $r_s$ by a strong increase of $\beta$ to get back the bouncing collapse behavior. We thus see that this
behavior is not a generic property of this model. In order to get such a behavior, one should choose parameters $M,\beta,r_s$ in such a way that $\beta\gg M$ and $r_S<0.3$, or if one wants to increase $r_s=0.4$, then $\beta>200$ should drastically increase as well. Otherwise, the system resembles the classical behavior as can be seen from figures \ref{fig:stationeryoutside branch  bounce} and \ref{fig:outside branch qm classic behavior}.

We now analyze the quantum corrected dynamics of the inside branch. The quantum corrected Hamiltonian of this branch, $H_- $, behaves very similarly to the outside branch. The only difference from the outside branch is that the asymptotic approach towards the horizon can happen from the inside. Indeed one might ask why these two branches of the quantum corrected Hamiltonian would behave so similarly despite the fact that their classical counterparts occupy completely different regions of space-time. Because the most drastic departure of the quantum corrected inside branch from the classical trajectories occur near $P_A=0$, where the classical Hamiltonian diverges, our regularization of equation \eqref{identity 2} might have caused this counterintuitive behavior. In order to see the bounce for the inside branch $\beta\gg M, r_s<0.3$, by increasing $r_s$ up to $r_s=0.4$ one can still get a bounce but one must drastically increase $\beta>200$. The coordinate radius $r_s$ being anything larger than the number we reported here force the system to behave like the classical portrait. Note that the $\beta$ that has appeared in our calculations in this section is a pure quantum parameter that was found using ACSQ. 

The value $P_A=0$ is of particular interest because for the bounce to occur the momentum needs to vanish; it is where the trajectory of the dust cloud changes from collapse to the expansion. Therefore we plot the phase-space portraits at $P_A=0$ for the outside branch as a function of $M$ for different $\beta\gg M$ to find \ref{fig:approximation rs 0.2}.

Investigating figure \ref{fig:approximation rs 0.2}, we see that the minimal area of the bouncing dust cloud lies outside of the photon sphere and the maximal area of recollapsing one between horizon and photon sphere. Considering that anything outside of the photon sphere is visible to the observer, this result is unfortunate, since it implies that the quantum corrected dynamics never exhibits anything even close to a black hole. Earlier we found an expression for the lifetime of the black hole from the point of view of the comoving observer (flat case). We also argued that the black hole is eternal for the closed case. This is, unfortunately, impossible to do for the stationary observer because when the bounce occurs the minimal radius is outside of the photon horizon $R=3M$. As seen by the stationary observer, the horizon never forms and the collapse never reaches a stage that resembles anything like a black hole. In a way, the bounce evades all conceptual problems by avoiding the horizon. Furthermore, the minimal area grows with increasing $\beta$ and $r_s<0.3$, while the maximal area decreases, suggesting that the quantum corrected dynamics strongly depends on fiducial vectors and thequantization ambiguity $\beta$. But interestingly, if one increases $r_s>0.5$, the quantum corrected dynamics will be extremely close to the classical one, which is the dashed green line in figure \ref{fig:stationeryoutside branch  bounce}. If we plot this for $P_A=0$ we find figure \ref{PA 0 rs 0.5}. Figure \ref{PA 0 rs 0.5} shows that all the content of the quantum corrected dynamics is outside of the horizon and therefore visible to the observer. But, unfortunately, there will be no bounce and recollapse in this case, see figure \ref{fig:classical rs big}. 

\section{switching between the comoving and the stationary observer}
\label{switching 10}
So far we have developed two quantum theories, one for a stationary observer and another for a comoving observer. We know that these two theories describe different sectors of space-time. If one wants to build up a full quantum theory of gravity from our construction one should find a way to include every observer and connect these two different theories, that is, to implement covariance. In this section we will try to find a way to switch between comoving and stationary observers. In order to do so, we closely follow the method described in \cite{tim}. For a comoving observer, we have
\begin{align}
\label{comoving switch 1}
     H=&-\frac{r_s^3}{6V_s} \frac{P_R^2}{R}-\frac{3V_s}{2r_s}kR,\\ 1=&\qty\bigg{R,P}
    . \label{comoving switch 2}
\end{align}
For a stationary observer, we have
\begin{align}
\label{stationery switch1}
    H=&-\frac{3V_S}{2p_k^{\frac{3}{2}}r_s^3}\left(p_k-1\right)R\\& \nonumber-\frac{3RV_s}{2p_k^{\frac{3}{2}}r_s^3}\tanh^2{(\frac{r_s^3\sqrt{p_k}P_R}{3V_s R })},\\1-\frac{\Pi^2}{R^2}=&\qty\bigg{R,\Pi}.
    \label{stationery switch2}
\end{align}
We consider the outside branch here, but the same can be done for the inside branch using the cotangent hyperbolic. By identifying $\Pi=R\tanh{(\frac{r_s^3\sqrt{p_k}}{3V_s }\frac{P_R}{R})}$, one observes that the Hamiltonian \eqref{stationery switch1} is identical to the comoving Hamiltonian \eqref{comoving switch 1} apart from a modified Poisson bracket, see \eqref{comoving switch 2} and \eqref{stationery switch2}. It reads 
\begin{equation}
     H=-\frac{3V_S}{2p_k^{\frac{3}{2}}r_s^3}\left(p_k-1\right)R-\frac{3V_s}{2p_k^{\frac{3}{2}}r_s^3}\frac{\Pi^2}{R}.
     \label{swithch station 3}
\end{equation}
In the following we want to show that this observation carries over from classical to quantum theory. 
Following what we did classically, where we promoted the function $\Pi(P,R)$ to the phase space coordinate $\Pi$, we proceed along the same way for their operators. Let us first consider the operator associated with $\Pi$ using the ACSQ map equation \ref{acsq} to find
\begin{equation}
    \begin{aligned}
        \Hat{\Pi}\psi(x)=&\frac{1}{2\pi\hbar c_{-1}^{\phi}}\int_0^{\infty} dR \int_{-\infty}^{\infty} dP\int_0^{\infty} dy \tanh{(\frac{r_s^3 \sqrt{p_k}}{3V_s}\frac{P}{R})}\\&e^{\frac{i}{\hbar}(x-y)P} \phi(\frac{x}{R})\phi^*(\frac{y}{R}) \psi(y),\\ \Hat{\Pi}\psi(x)=&\frac{1}{2\pi \hbar c_{-1}^{\phi}}\int_0^{\infty}dR \int_{-R}^{R} \frac{d\Pi}{1-\frac{\Pi^2}{R^2}}\int_0^{\infty} dy \,\frac{\Pi}{R}\\&e^{\frac{i}{\hbar}(x-y)\frac{R}{2}\ln{(\frac{1+\frac{\Pi}{R}}{1-\frac{\Pi}{R}}})}\phi(\frac{x}{R})\phi^*(\frac{y}{R}) \psi(y).
    \end{aligned}
    \label{acsq Pi}
\end{equation}
Equation \eqref{acsq Pi} shows that using ACSQ maps we have
\begin{equation}
    \begin{aligned}
        \Hat{f}\psi(x)&=\frac{1}{2\pi\hbar c_{-1}^{\phi}}\int_0^{\infty} \frac{dR}{R}\int_{-R}^{R}\frac{d\Pi}{\abs{1-\frac{\Pi^2}{R^2}}}\int_0^{\infty} dy \,f(R,\Pi)\\&e^{\frac{i}{\hbar}(x-y)\frac{R}{2}\ln{\abs{\frac{1+\frac{\Pi}{R}}{1-\frac{\Pi}{R}}}}}\phi(\frac{x}{R})\phi^*(\frac{y}{R}) \psi(y),
    \end{aligned}
    \label{acsq pi finale 6}
\end{equation}
where we have used 
\begin{align*}
    \tanh^{-1}{x}=&\ln{\frac{1+x}{1-x}},\\
    \tanh{(\frac{r_s^3\sqrt{p_k}}{3V_s }\frac{P_R}{R})}=&\frac{\Pi}{R}.
    \end{align*}
    Furthermore, equation \eqref{acsq pi finale 6} assigns $\Hat{f}$ to the phase-space function $f(\Pi,R)$, and it is indeed equivalent to the ACSQ map with alternative half line parametrization 
\begin{align*}
    R\in \mathbb{R}_+&, \frac{R}{2}\ln{\abs{\frac{1+\frac{\Pi}{R}}{1-\frac{\Pi}{R}}}}\in \mathbb{R}.
\end{align*}
This would also imply a new coherent state which reads
\begin{equation}
    \bra{x}\ket{\Pi,R}=\frac{1}{\sqrt{R}}e^{\frac{i}{\hbar}\frac{R}{2}\ln{\abs{\frac{1+\frac{\Pi}{R}}{1-\frac{\Pi}{R}}}}x} \psi(\frac{x}{R}).
\end{equation}
Note that these states lead to the resolution of identity; see \cite{gozdz2021dependence} for more details.

Let us now show that these quantization maps indeed reproduce \eqref{stationery switch2}. We know from ACSQ how $\hat{R}$ looks like, see \eqref{R Finale}; therefore we can find the commutator
\begin{equation}
    \begin{aligned}
        \left[\Hat{R},\Hat{\Pi}\right]=&\frac{c_0^{\phi}}{2\pi \hbar (c_{-1}^{\phi})^2}\int_0^\infty \frac{dR}{R}\int_{-R}^{R} \frac{d\Pi}{1-\frac{\Pi^2}{R^2}}\int_0^{\infty} dy\,(x-y) \\&\Pi \,e^{\frac{i}{\hbar}(x-y)\frac{R}{2}\ln{\abs{\frac{1+\frac{\Pi}{R}}{1-\frac{\Pi}{R}}}}} \phi(\frac{x}{R})\phi^*(\frac{y}{R}) \psi(y)\\&=-\frac{ic_0^{\phi}}{2\pi(c_{-1}^{\phi})^2}\int_0^{\infty} \frac{dR}{R}\int_{-R}^{R} d\Pi\\ &\int_0^{\infty} dy\, \textrm{sgn}\left(1-\frac{\Pi^2}{R^2}\right) \Pi
        \pdv{\Pi} e^{i(x-y)\frac{R}{2}\ln{\abs{\frac{1+\frac{\Pi}{R}}{1-\frac{\Pi}{R}}}}}\\&\phi(\frac{x}{R})\,\phi^*(\frac{y}{R}) \psi(y).
        \end{aligned}
\end{equation}
Applying integration by part to the last term, we finally find
\begin{equation}
\begin{aligned}
  \left[\Hat{R},{\Hat{\Pi}}\right] =& \frac{ic_0^{\phi}}{2 \pi( c_{-1}^{\phi})^2}\int_0^{\infty} \frac{dR}{R}\int_{-R}^{R} \frac{d\Pi}{\abs{1-\frac{\Pi^2}{R^2}}}\int_0^{\infty} dy \left(1-\frac{\Pi^2}{R^2}\right),\\&e^{\frac{i}{\hbar}(x-y)\frac{R}{2}\ln{\abs{\frac{1+\frac{\Pi}{R}}{1-\frac{\Pi}{R}}}}} \phi(\frac{x}{R})\phi^{*}(\frac{y}{R}) \psi(y)\\&\left[\Hat{R},\Hat{\Pi}\right]=i\hbar \frac{c_0^{\phi}}{c_{-1}^{\phi}}{\widehat{\qty\bigg{{R,\Pi}}}}.
  \end{aligned}
\end{equation}

We have thus shown that the quantization maps indeed reproduce \eqref{stationery switch2} with a difference of a coefficient, but this is not a problem because one can choose fiducial vectors in a certain way or rescale $\Hat{R},\Hat{\Pi}$ and make the additional constant disappear. Furthermore,  we have demonstrated here that the quantum theories for comoving observer and stationary observer can be related by modifying Poisson brackets. In ACSQ this is equivalent to switching between parametrizations of the half line (affine group). But this is still at the classical level and so our next task is to investigate whether it is possible to discuss relation between comoving and stationary observer purely at the quantum level. In order to do so, let us consider what we did so far with the map $\tau$. If $\tau$ is a map at the quantum level, it has to be linear and bijective, at least when it is restricted to operators corresponding to phase space functions. But it cannot be an isomorphism, since it changes the commutation relations, but leaves the identity invariant,
\begin{equation}
    \tau(\left[\Hat{R},\Hat{P}\right])=i\hbar \tau(\mathbb{I})=i\hbar \mathbb{I}\neq \left[\tau(\hat{R}),\tau(\hat{P})\right]=\left[\Hat{R},\Hat{\Pi}\right].
\end{equation}

If we follow the reasoning of \cite{tim} and formulate this map, we find out that it has to be linear. Since linear operators have adjoints, we find that $\tau$ must be unitary. But this is impossible here because it leads to a contradiction and implies that $\tau$ has to be an isomorphism of an operator algebra; for more details, see \cite{gozdz2021dependence}. We conclude that a switch between these two observers cannot happen at the level of the Hilbert space. The switch is also not unitary, but it can be understood at the level of operator algebra, namely changing the parametrization of the half line in the ACSQ scheme. This is undesirable because one expects the states in full quantum gravity to be connected via unitary transformation. But one can defend our analysis here by claiming that both observers see the bounce, so our model is consistent.

%%%%%%%%%%%%%%%%%%%%%%%%%%%%%%%%%%%%%%%%%%%
\section{Summary and conclusion}
\label{summary}

In our paper we have filled an important gap in the literature on the quantum Oppenheimer-Snyder (OS) model in geometrodynamics variables. So far, quantization has addressed the flat sections of the dust cloud. But the covariance of the classical theory demands that all possible foliations be considered, which includes the cases of positive and negative curvature. This was achieved here.

We have introduced the canonical OS-model and applied the Kucha\v{r} decomposition. Thereby the constraints were greatly simplified, but a non-canonical boundary term appeared in the action. In the flat case, this boundary term can be written as a total derivative \cite{tim1}. In the general case, one has to use a different approach. We have employed generalized Painlev\'e-Gullstrand coordinates for the open case and Gautreau-Hoffman coordinates for the closed case and have promoted the coordinate transformation between dust proper time and Schwarzschild Killing time to a canonical transformation. 
We have derived the general Hamiltonian for a stationary observer and shown that in the flat case one can recover from it the Hamiltonian presented in \cite{tim1}.  

Due to the complexity of the quantization for the non-marginal case,
we were unable to perform the standard Dirac quantization. Instead, we
have successfully applied Affine Coherent State Quantization
(ACSQ). With this, every semibounded phase space function can be
mapped to a self-adjoint operator. We have found that at least for a
certain parameter range the classical singularity is avoided in the
sense that there is a bounce and recollapse of wave packets as seen by
a stationary observer. This is reminiscent of the case studied earlier
for collapsing {\em shells} \cite{Hajicek:2001yd,yeom2022}. We have
also investigated the switch to a 
comoving observer and found consistency. 

By quantizing the comoving observer for the closed case $(k=1)$ in the
ACSQ scheme we have found that there is a bounce that is accompanied
by a recollapse. The system experiences an oscillatory behavior
between bounce and recollapse, see figure~\ref{fig:comoving
  closed}. Furthermore, we have established that these oscillations
are located well within the horizon, so the dynamics of the comoving
observer suggests that we deal with an eternal black hole. This seems
to solve the lifetime problem at least from the point of view of the
comoving observer, as an eternal lifetime is certainly not
short. Unfortunately, the same construction is impossible to implement
for a stationary observer. The reason is that when the bounce occurs,
the minimal radius is located outside the photon sphere $R=3M$ and the
maximal radius of the recollapse between horizon and photon sphere
(note that anything outside the photon sphere is visible to the
observer); see figure \ref{fig:approximation rs 0.2} . As seen by the
stationary observer, the horizon never forms, and the collapse never
reaches a stage that resembles anything like a black hole. Hence it is
impossible to define a lifetime from the point of view of the
stationary observer. We recall that the OS-model exterior is of
Schwarzschild and the interior of FLRW form. One should analyze these
two regions and match them together using the junction conditions
given, for example, in \cite{Poisson:2009pwt}. Because we have fully
deparametrized the exterior at the classical level, a matching of this
form will not survive the quantization. One reasonable approach to
address this issue is to take the quantum corrections that we derived
here and formulate them as effective matter contributions. One can
then investigate what exterior can be matched to it,
cf. \cite{achour2020bouncing2}, \cite{Achour2020BouncingCO1}. In this
context it would also be of interest to establish a connection to
heuristic investigations arriving at a discrete spectrum for the
quantum-gravitational wave function (see
e.g. \cite{casadio2022}). Finally, a most
important issue is to include Hawking radiation and to study the
information-loss problem. 
We leave these and other issues for future publications.

%%%%%%%%%%%%%%%%%%%%%%%%%%%%%%%%%%%%%%%%%%%%%%%%%%%%%%%%%

\section*{Acknowledgments}
We thank Domenico Giulini, Tim Schmitz, and Enes Akta\c{s} for 
fruitful and inspiring discussions and critical comments.

%%%%%%%%%%%%%%%%%%%%%%%%%%%%%%%%%%%%%%%%%%%%%%%%%%%%%%%%%
\appendix
\label{sec:app_quantization acsq}
\section{Quantizing phase space functions $P, R, RP,P^2,R^{\beta}$ with ACSQ}

Here we derive the ACSQ version of elementary phase space functions which we will heavily use in the text. We have
\begin{equation}
\begin{aligned}
    \Hat{R}\psi(x)=&\frac{1}{2 \pi h\, c_{-1}^{\phi}}\int_{0}^{\infty} dR \int_{-\infty}^{\infty} dP \\ &\int_0^{\infty} dy\, e^{\frac{i}{h}(x-y)p} \phi(\frac{x}{R})\,     \phi^*(\frac{x}{R}) \psi(y).
    \end{aligned}
\end{equation}
Using the identities 
\begin{align}
\label{delta identity 1}
    \int_{-\infty}^{\infty} dP\, e^{\frac{i}{h}(x-y)P}&=2\pi \, \delta(x-y)\\
    \int_{-\infty}^{\infty} dP \,P \, e^{\frac{i}{h}(x-y)P}&=-2\pi i \, \delta'(x-y)
    \label{delta identity 2}
\end{align}
we get 
\begin{equation}
    \Hat{R}\psi(x)= \frac{1}{c_{-1}^{\phi}} \int_{0}^{\infty} dR \abs{\phi(\frac{x}{R})}^2 \psi(x).
    \label{R acsq intermediate}
\end{equation}

By changing the variables $\frac{x}{R}=u$ we further modify equation \eqref{R acsq intermediate}
to find 
\begin{equation}
\Hat{R}\psi(x)=\frac{1}{c_{-1}^{\phi}}\int_{0}^{\infty} \frac{du}{u^2} x\psi(x)=\frac{c_0^{\phi}}{c_{-1}^{\phi}} x\, \psi(x).
\label{R Finale}
\end{equation}
Next we calculate $R^{\beta}$, $\beta \in \mathbb{N}$, which generalizes the equation we get for $R$,
\begin{equation}
\begin{aligned}
    \Hat{R}^{\beta} \psi(x)=&\int_{-\infty}^{\infty} dP \int_{0}^{\infty} dR \\&\int_0^{\infty}dy \, \frac{R^{\beta}}{R} e^{ip(x-y)}\phi(\frac{x}{R})  \phi(\frac{y}{R}) \psi(y).
    \end{aligned}
\end{equation}
By a change of variable $\frac{x}{R}=u$ we get 
\begin{equation}
   \int_0^{\infty} \left(\frac{x}{u}\right)^{\beta+1} \frac{du}{x}\abs{\phi(u)}^2\psi(x).
   \label{R^beta intermediate}
\end{equation}
Further simplifying equation \eqref{R^beta intermediate} \eqref{fiducial1} 
we get 
\begin{equation}
    \Hat{R}^{\beta}=\frac{c_{\beta-1}^{\phi}}{c_{-1}^{\phi}}R^{\beta}\psi(x).
    \label{R^beta finale}
\end{equation}
Note that for $\beta=1$ one recovers equation \eqref{R Finale}. Moreover, let us calculate $\Hat{P}$ to find
\begin{align}
    \Hat{P}\psi&=\frac{1}{2\pi h \, c_{-1}^{\phi}}\int_{0}^{\infty} dR \int_{-\infty}^{\infty} dP \int_{0}^{\infty} \, \frac{P}{R} \,e^{\frac{i}{h}P(x-y)} \\ \nonumber &\phi(\frac{x}{R})\phi^*(\frac{x}{R}) \psi(y) dy\\
    &=\frac{-ih}{c_{-1}^{\phi}}\int_0^{\infty} \frac{dR}{R}\phi(\frac{x}{R})\pderivative{x} \phi^*(\frac{x}{R}) \psi(x)\\&=-ih\psi'(x)-ih\gamma \frac{\psi(x)}{x}\label{acsq p final}
    \\ \gamma&=\frac{\int_0^{\infty}du\, \phi(u)\phi^*(u)'}{c_{-1}^{\phi}}.
    \end{align}
For real fiducial vectors $\gamma=0$ with $u=\frac{x}{R}$. Furthermore, we calculate the dilation operator $D=RP$ to find
\begin{align}
   \Hat{R} \Hat{P}\psi(y)&=\frac{1}{2\pi h\, c_{-1}^{\phi}}\int_0^{\infty} dR\\& \nonumber \int_{-\infty}^{\infty}dP \int_0^{\infty} dy\, P\, e^{\frac{i}{h}P(x-y)}
    \phi(\frac{x}{R})\phi^*(\frac{x}{R}) \psi(y)\\
    &=-\frac{ih}{c_{-1}^{\phi}}\int_0^{\infty} dR\, \phi(\frac{x}{R})\pderivative{x}\phi^*(\frac{x}{R}) \psi(y)\\
    &=-ih \frac{c_0^{\phi}}{c_{-1}^{\phi}} x \psi'(x)-\frac{ih\,\psi(x)}{c_{-1}^{\phi}}\int_0^{\infty}\frac{du}{u}\phi(u) \phi^*(u)'.
    \label{dilation finale}
\end{align}
Constants on the right-hand side can be collected to form a new constant called $\lambda$ with $u=\frac{x}{R}$,
\begin{equation}
\lambda=\frac{\int_0^{\infty}\frac{du}{u}\, \phi(u)\phi^*(u)'}{c_{-1}^{\phi}}.
\label{acsq lambda}
\end{equation}
Note that fiducial vector $\phi$ comprises a family of well behaved functions which are fixed and arbitrary. Therefore one can simplify the numerator in equation \eqref{acsq lambda} by remembering the fact that $\phi \phi'=\frac{1}{2}\dv{u}(\phi^2)$(prime denotes differentiation with respect to u); employing integration by parts and making use of \eqref{fiducial1}, we find for real fiducial vectors 
\begin{equation}
    \lambda =\frac{c_0^{\phi}}{2\,c_{-1}^{\phi}}.
\end{equation}
Note that integration by part does not lead to boundary terms, since the fiducial vectors are chosen in a way to vanish at zero and infinity. Last but not least we perform ACSQ quantization for ${P}^2$. 

Starting with the definition \eqref{acsq} we find 
\begin{equation}
\begin{aligned}
    \Hat{P}^2\psi(y)=&\frac{1}{2\pi h \, c_{-1}^{\phi}}\int_{0}^{\infty} dR\\& \int_{-\infty}^{\infty} dP \int_{0}^{\infty} \, \frac{P^2}{R} \, \,e^{\frac{i}{h}P(x-y)} \phi(\frac{x}{R})\phi^*(\frac{x}{R}) \psi(y) dy.
    \label{P^2 intermediate first}
    \end{aligned}
\end{equation}
Using the fact that 
\begin{equation*}
    \int_{-\infty}^{\infty}P^2 \,e^{\frac{i}{h}P(x-y)}=2\pi i \, \delta''(x-y)
\end{equation*}
we further simplify equation \eqref{P^2 intermediate first} to get 
\begin{equation}
    \Hat{P}^2 \psi(y)=\frac{1}{2\pi h\, c_{-1}^{\phi}} \int_{0}^{\infty} dR \frac{dR}{R} \phi(\frac{x}{R})\pdv[2]{x}\left(\phi^*(\frac{x}{R})\psi(y)\right)
.\end{equation}
After computing derivatives and making some straightforward calculations we get
\begin{equation}
\begin{aligned}
  \Hat{P}^2\psi(y)=&\int_0^{\infty}\frac{1}{R^3} \phi(\frac{x}{R})\phi''(\frac{x}{R})\psi(x) dR\\&+\int_{0}^{\infty}\frac{2}{R^2}\phi(\frac{x}{R})\phi'(\frac{x}{R}) \psi(x) dR\\&+\int_{0}^{\infty} \phi(\frac{x}{R})\phi(\frac{x}{R})\psi''(x) dR.
  \label{P^2 intermediate}
  \end{aligned}
\end{equation}
By changing the variable $\frac{x}{R}=u$ and $du=-\frac{x}{R^2}dR$ and simplifying equation \eqref{P^2 intermediate} we get 
\begin{align}
\label{P^2 intermediate1}
   \hat{P}^2 \psi(x)=&\frac{1}{\, c_{-1}^{\phi}} \int_0^{\infty} \psi''(x) \phi^2(u) \frac{du}{u}\\-&\nonumber\frac{2}{ \, c_{-1}^{\phi}}\int_{0}^{\infty} \frac{du}{x} \phi(u) \phi'(u) \psi'(x)\\-&\int_{0}^{\infty} \frac{du}{x}\phi(u)\phi''(u) \frac{u}{x} \psi(x). \nonumber
   \end{align}
By using equation \eqref{fiducial1}, the first term in equation \eqref{P^2 intermediate1} can be readily written with respect to $c_{\alpha}^{\phi}$,
\begin{equation}
    \frac{\Hat{P}^2 c_{-1}^{\phi}}{c_{-1}^{\phi}}.
    \label{P^2 intermediate 2}
\end{equation}
Using integration by parts and remembering that fiducial vectors $\phi$ vanish at boundaries, and moreover using $\phi \phi'= \frac{1}{2}\dv{u}\left(\phi^2\right)$ and evaluating integration by part, the second term of equation \eqref{P^2 intermediate1} vanishes. For the third term we use $\phi \phi''=\dv{u}\left(\phi \phi'\right)-\phi'^2$ and make a partial integration to get
\begin{equation}
\frac{1}{c_{-1}^{\phi}}\int_0^{\infty} \frac{du}{x^2} u\, \phi'(u)^2 \psi(x).
\label{P^2 intermediate 3}
\end{equation}
Adding equations \eqref{P^2 intermediate 2} \eqref{P^2 intermediate 3} we get $\Hat{P}^2$,
\begin{align}
   \Hat{P}^2=P^2+\frac{K_{\phi}}{x^2},
   \label{P^2 finale}
\end{align}
with $K_{\phi}=\int_0^{\infty}\phi'^2(u) u\, \frac{du}{c_{-1}^{\phi}}$.

\section{ACSQ lower symbol calculations}\label{sec:app_lower symbol calculation}
As discussed in section \ref{affine covariant integral quantization}, lower symbols are used for a
semiclassical analysis of the system. In the text, we need the lower symbol versions of some elementary phase space functions; here we present the details of the calculations.
We start with the lower symbol of the position operator using \eqref{lower symbol} to find 
\begin{align}
    \Check{R}=\frac{c_0^{\phi}}{c_{-1}^{\phi}}\int_0^{\infty} \frac{dx}{R} x\abs{\phi(\frac{x}{R})}^2=\frac{c_{-3}^{\phi}c_0^{\phi}}{c_{-1}^{\phi}}R.
    {\label{lower symbol R}}
\end{align}
Next we calculate the lower symbols of $P$,
\begin{align}
\label{lower symbol P}
    \Check{P}&=-ih \int_0^{\infty} \frac{dx}{R}e^{-\frac{i}{h}px}\phi^*(\frac{x}{R})\left(\pdv{x}+\frac{\gamma}{x}\right) e^{\frac{i}{h}px}\phi(\frac{x}{R})\\
    \Check{P}&=P-\frac{ih}{R}\int_0^{\infty} du \left(\phi(u)^2\right)'\\
    u&=\frac{x}{R}. \nonumber
\end{align}
as well as the lower symbols of $R^{\beta}$ and $\beta \in \mathbb{N}$,
\begin{align}
   \Check{R}^{\beta} =\frac{c_{\beta -1}}{c_{-1}}\int_0^{\infty} \frac{dx}{R} x^{\beta}\, \abs{\phi(\frac{x}{R})}^2 dx.
   \label{lower symbol R^ beta}
\end{align}
With $u=\frac{x}{R}$ , $dx=Rdu$ we get 
\begin{align}
   \Check{R}^{\beta}=\frac{c_{\beta -1}}{c_{-1}}\int_0^{\infty} \frac{dx}{R} (Ru)^{\beta}\, \abs{\phi(u)}^2 Rdu.
   \label{lower symbol R beta intermediate}
\end{align}
By comparing \eqref{lower symbol R beta intermediate} with \eqref{fiducial1} we arrive at
\begin{equation}
    \Check{R}^{\beta}=\frac{c_{-\beta-2}^{\phi}\,c_{\beta-1}^{\phi}}{c_{-1}^{\phi}}R^{\beta}.
    \label{lower symbol r^beta finale}
\end{equation}
Note that $\beta=1$ reproduces \eqref{lower symbol R}. Last but not least let us calculate the lower symbol for $\check{P}^2$,
\begin{align}
    \check{P}^2=\int_0^{\infty} \frac{dx}{R}\: e^{-\frac{i}{h}px}\,  \phi(\frac{x}{R})\left(\pdv[2]{x}+\frac{c_{-3}^{\phi}}{c_{-1}^{\phi}x^2}\right)\phi(\frac{x}{R})e^{\frac{i}{h}px}.
\end{align}
Making the change of variables $u=\frac{x}{R}$ and evaluating integrals we get after some long but straightforward calculations for the lower symbol of $\Check{P}^2$ the expression
\begin{align}
    \check{P}^2&=P^2+\frac{c(\phi)}{R^2}\\
    c(\phi)&=\int_0^{\infty} \phi'^2(u)\left(1+u\frac{c_0^{\phi}}{c_{-1}^{\phi}}\right)du.
    \label{lower symbol P^2 finale}
\end{align}

\section{ACSQ calculation for comoving OS observer}
\label{ACSQ calculation for comoving OS observer}

As we have found in equation \eqref{rescaled hamiltonian}, the Hamiltonian for comoving observer in general case (flat and curved spaces) after deparametrization with respect to dust proper time  reads
\begin{equation}
    H=-\frac{r_s^3}{6V_s}\frac{P_R^2}{R}-\frac{3V_sk}{2r_s}R,
    \label{comoving hamil append.1}
\end{equation}
which $k$ is the curvature parameter and assumes values ($0,\pm 1$). Here we would like to briefly discuss how one can quantize above Hamiltonian with the ACSQ scheme. We found the ACSQ quantized version of the second term in equation \eqref{comoving hamil append.1} to be like \eqref{R Finale}. Moreover, for the lower symbol of this term we have \eqref{lower symbol R}. Therefore we can find the lower symbol of the second term by taking \eqref{lower symbol R} and multiply it by $\frac{3V_s}{2r_s}k$. Therefore our task now is to find the ACSQ version of the first term in \eqref{comoving hamil append.1} and subsequently the lower symbol of it.  Starting with the definition given in \eqref{acsq}, we have
\begin{align}
    \Hat{H}=\frac{1}{2\pi \hbar c_{-1}^{\phi}}\int_0^{\infty} dR\int_{-\infty}^{\infty} dP \, H(P,R)\ket{P,R}\bra{P,R}.
\end{align}
We insert ${H}=-\frac{P^2}{2R}$ in the above definition to get
\begin{equation}
    \begin{aligned}
      \Hat{H}\psi(x)&=\frac{1}{2\pi\hbar c_{-1}^{\phi}}\int_0^{\infty}dR\int_{-\infty}^{\infty} dP \\& \nonumber\int_0^{\infty} dy \frac{P^2}{R^2} e^{\frac{i}{\hbar}\left(x-y\right)P} \phi(\frac{x}{R})\phi^*(\frac{x}{R})\psi(y)\\&=\frac{\hbar^2}{2c_{-1}^{\phi}}\int_0^{\infty} \frac{dR}{R^2} \phi(\frac{x}{R})\pdv[2]{x}\phi^*(\frac{x}{R}) \psi(y),
    \end{aligned}
\end{equation}
where we have used the identity  $\int_{-\infty}^{\infty}dP P^2e^{i(x-y)P}=-2\pi\delta''(x-y)$, and we only use real fiducial vectors. Calculating the derivatives we have 
\begin{equation}
    \begin{aligned}
      \Hat{H}\psi(x)&=\frac{\hbar^2}{2c_{-1}^{\phi}}\int_0^{\infty}dR\,\phi(\frac{x}{R})\\ \nonumber&\pdv{x}\left(\frac{1}{R}\phi'(\frac{x}{R})\psi(y)+\phi(\frac{x}{R}) \psi'(y))\right)\\
      &=\frac{\hbar^2}{2c_{-1}^{\phi}}\int_0^{\infty}\frac{dR}{R^4}\,\phi(\frac{x}{R})\phi''(\frac{x}{R})+\\ & \nonumber\frac{\hbar^2}{c_{-1}^{\phi}}\int_0^{\infty}\frac{dR}{R^3}\,\phi(\frac{x}{R}) \phi'(\frac{x}{R})+\\&\frac{\hbar^2}{2c_{-1}^{\phi}}\int_0^{\infty}\frac{dR}{R^2}\phi(\frac{x}{R})\phi(\frac{x}{R})\psi''(y).
    \end{aligned}
\end{equation}
Now we employ the change of variable $u=\frac{x}{R}$ to get 
\begin{align}
\label{os acsq intermediate}
    \Hat{H}\psi(y)=&\frac{\hbar^2}{2c_{-1}^{\phi}}\int_0^{\infty}\frac{du}{x^3}\,u^2 \phi(u)\phi''(u)\psi(y)\\&-\frac{\hbar^2}{c_{-1}^{\phi}}\int_0^{\infty} \frac{du}{x^2} \phi(u)\phi'(u)\psi(y) \\& \nonumber+\frac{\hbar^2}{2c_{-1}^{\phi}} \int_0^{\infty} \frac{du}{x}\phi(\frac{x}{R}) \phi(\frac{x}{R})\psi''(y).\nonumber
\end{align}
Using the following identities for fiducial vectors,
\begin{align}
\label{phi identity}
    \phi(u)\phi''(u)&=\dv{u}\left(\phi(u)\phi'(u)\right)-\phi'(u)^2\\
    \phi(u)\phi'(u)&=\frac{1}{2}\dv{u}\left(\phi(u)^2\right),
\end{align}
evaluating integration by parts and employing equation \eqref{fiducial1} in order to turn $du$ integrations to constants $c_{\alpha}^{\phi}$, we find for the first term 
\begin{equation}
    \begin{aligned}
      &\int_0^{\infty} \frac{du}{x^3} u^2 \dv{u}\left(\phi(u)\phi'(u)-\phi'(u)^2\right)\psi(y)\\&=\int_0^{\infty}\frac{du}{x^3}\dv{u}(\phi(u)\phi'(u))u^2\psi(y)-\int_0^{\infty}u^2 \frac{\phi'(u)^2\psi(y)}{x^3} du.
      \label{dog acsq}
    \end{aligned}
\end{equation}
After evaluating integration by parts twice for the first term we find out that $du$-integration gives one, while the second term is $c_{-4}^{\phi}$. Therefore \eqref{dog acsq} using \eqref{fiducial1} can be simplified to 
\begin{equation*}
    \frac{1-c_{-4}^{\phi'}}{x^3}.
\end{equation*}
The same thing can be done for the second and the third term in \eqref{os acsq intermediate},
and we finally get 
\begin{equation}
   \Hat{H} \psi=\frac{\hbar^2}{2c_{-1}^{\phi}}\left(\frac{1-c_{-4}^{\phi'}}{x^3}\psi(x)-\frac{1}{x^2}\psi'(x)+\frac{1}{x}\psi''(x)\right).
   \label{acsq os finale}
\end{equation}

\section{Lower symbol calculation for comoving observers}
\label{lower symbol calculation for comoving os model}
In this section we calculate the lower symbols of ${H}=-\frac{P^2}{2R}$. Starting from the definition \eqref{lower symbol 1}, we find 
\begin{equation}
\begin{aligned}
    \check{H}=\frac{\hbar^2}{2Rc_{-1}^{\phi}}&\int_0^{\infty} dx\, e^{-\frac{i}{\hbar}p.x} \phi(\frac{x}{R})\\&\left(\frac{1-c_{-4}^{\phi}}{x^3}-\frac{1}{x^2}\pdv{x}+\frac{1}{x}\pdv[2]{x}\right)e^{\frac{i}{\hbar}p.x} \phi(\frac{x}{R}).
    \end{aligned}
    \label{lower symbol comoving intermediate}
\end{equation}
Note that all fiducial vectors are taken to be real in our calculation. We can, of course, calculate the lower symbols directly by using \eqref{lower symbol}. But because the ACSQ version of $-\frac{P^2}{2R}$ is readily available from the last section \ref{ACSQ calculation for comoving OS observer}, using the ACSQ version means we only need to calculate one integral. For the first term of the right-hand side \eqref{lower symbol comoving intermediate} we have
\begin{align}
  \textit{first term}  =&\frac{\hbar^2}{2c_{-1}^{\phi}} \int_0^{\infty} dx \: e^{-\frac{i}{h}p.x} \phi(\frac{x}{R}) \left(\frac{1-c_{-4}^{\phi'}}{x^3}\right)e^{\frac{i}{h}p.x}\phi(\frac{x}{R})\\
  =& \frac{\hbar^2}{2c_{-1}^{\phi}}\int_0^{\infty} \frac{dx}{R} \, \frac{\abs{\phi(\frac{x}{R})}^2}{x^3}\\& \nonumber -\frac{\hbar^2}{2c_{-1}^{\phi}}\int_0^{\infty}\frac{dx}{R}\left(\frac{c_{-4}^{\phi'}}{x^3}\right)\abs{\phi(\frac{x}{R})}^2.
\end{align}
Employing the change of variables $u=\frac{x}{R}$ and using \eqref{fiducial1} we get
\begin{align}
    \textit{first term}=&\frac{\hbar^2}{2c_{-1}^{\phi}}\int_0^{\infty} du\, \frac{\phi(u)^2}{R^3u^3}\\& \nonumber-\frac{\hbar^2}{2c_{-1}^{\phi}}\int_0^{\infty} du \frac{c_{-4}^{\phi'}}{R^3u^3} \phi(u)^2 \\&=\frac{\hbar^2c_{1}^{\phi}}{2c_{-1}^{\phi}R^3}-\frac{\hbar^2 c_{1}^{\phi} c_{-4}^{\phi'}}{2c_{-1}^{\phi}R^3}.
\end{align}
Next we evaluate the second term on the right-hand side of \eqref{lower symbol comoving intermediate}. We find 
\begin{align}
   \textit{second term}=& \frac{\hbar^2}{2c_{-1}^{\phi}}\int_0^{\infty} e^{-\frac{i}{\hbar}p.x} \,\phi(\frac{x}{R}) \left(-\frac{1}{x^2} \pdv{x}\right) e^{\frac{i}{\hbar}p.x} \phi(\frac{x}{R})\\
   =& -\frac{i\hbar}{2c_{-1}^{\phi}}\int_0^{\infty}\frac{dx}{R\,x^2}P\abs{\phi(\frac{x}{R})}^2\\&- \nonumber\frac{\hbar^2}{2c_{-1}^{\phi}}\int_0^{\infty} \frac{dx}{R^2x^2} \phi(\frac{x}{R})\phi'(\frac{x}{R}) .
\end{align}
We apply the change of variable $u=\frac{x}{R}$, and use equation \eqref{fiducial1} to find 
\begin{align}
   \textit{second term}=& -\frac{i\hbar}{2c_{-1}^{\phi}}\int_0^{\infty}\frac{du}{R^2u^2} P \phi(u)^2\\ -\nonumber& \frac{\hbar^2}{2c_{-1}^{\phi}}\int_0^{\infty}\frac{du}{R^3u^2} \phi(u)^2\\=&-\frac{i\hbar P c_0^{\phi}}{2c_{-1}^{\phi}R^2}-\frac{\hbar^2 c_1^{\phi}}{2c_{-1}^{\phi}R^3}.
\end{align}
We calculate the third term of equation \eqref{lower symbol comoving intermediate} to get
\begin{align}
    \textit{third term}=&\frac{\hbar^2}{2c_{-1}^{\phi}}\int_0^{\infty} dx\, e^{-\frac{i}{\hbar}p.x} \phi(\frac{x}{R})\left(\frac{1}{x}\pdv[2]{x}\right)e^{\frac{i}{\hbar}p.x}\phi(\frac{x}{R})\\=& -\frac{1}{2c_{-1}^{\phi}}\int_0^{\infty} \frac{dx}{R\,x}\abs{\phi(\frac{x}{R})}^2 P^2\\& \nonumber +\frac{i\hbar}{c_{-1}^{\phi}}\int_0^{\infty}\frac{dx\,P}{R^2x}\phi(\frac{x}{R})\phi'(\frac{x}{R})\\& \nonumber+ \frac{\hbar^2}{2c_{-1}^{\phi}}\int_0^{\infty} dx\,\frac{\phi(\frac{x}{R})\phi''(\frac{x}{R})}{R^3x}.\nonumber
\end{align}
We employ the change of variable $\frac{x}{R}=u$ and use equation \eqref{fiducial1} to get
\begin{align*}
  \textit{third term}=& -\frac{1}{2c_{-1}^{\phi}} \int_0^{\infty} du \frac{\phi(u)^2}{u}\frac{P^2}{R} \\& \nonumber +\frac{i\hbar}{c_{-1}^{\phi}}\int_0^{\infty} du \frac{P}{R^3u} \dv{u}\left(\phi(u)\right)^2\\& + \frac{\hbar^2}{2c_{-1}^{\phi}}\int_0^{\infty} \frac{du}{R^3u}\phi(u)\phi''(u).
\end{align*}
The first two terms on the right-hand side of the last equation can readily be written with respect to $c_{\alpha}^{\phi}$ using equation \eqref{fiducial1}. The last term though requires more work. In order to simplify it we use the identity \eqref{phi identity} to get 
\begin{align}
    =&\frac{\hbar^2}{2c_{-1}^{\phi}} \int_0^{\infty} \frac{du}{R^3u} \left[\dv{u}\left(\phi(u)\phi'(u)\right)-\phi'(u)^2\right]\\
    =&- \frac{\hbar^2}{2c_{-1}^{\phi}}\int_0^{\infty}\frac{du}{R^3u^2}\dv{u}\left(\phi(u)^2\right)\\& \nonumber-\frac{\hbar^2}{2c_{-1}^{\phi}R^3}\int_0^{\infty}\frac{\phi'(u)^2}{u}du\\=&\frac{\hbar^2c_1^{\phi}}{2c_{-1}^{\phi}R^3}-\frac{\hbar^2c_{-1}^{\phi'}}{2c_{-1}^{\phi}R^3}.
\end{align}
We thus find for the third term 
\begin{equation}
    \textit{third term}=-\frac{P^2}{2R}+\frac{i\hbar c_0^{\phi}P}{2c_{-1}^{\phi}R^2}-\frac{\hbar^2c^1_{\phi}}{2c_{-1}^{\phi}R^3}.
\end{equation}
Taking everything together, we arrive at the lower symbol for the comoving observer,
\begin{equation}
  \check{H}=-\frac{P^2}{2R}-\hbar^2\frac{ c_{-1}^{\phi'}+c_{1}^{\phi}\left(c_{-4}^{\phi'}-1\right)}{2R^3c_{-1}^{\phi}}. 
  \label{lower symbol comoving os finale app}
\end{equation}
\clearpage

%%%%%%%%%%%%%%%%%%%%%%%%%%%%%%%%%%%%%%%%%%%%%%%%%%%%%%%%%%%%%%%%%%

%%%%%%%%%%%%%%%%%%%%%%%%%%%%%%%%%%%%%%%%%%%%%%%%%%%%%%%%%%%%%%%%%

\end{document}